\documentclass[12pt]{article}
\usepackage{epsfig}
\usepackage{axodraw}
 
\newcommand{\as}{\alpha_{\mathrm{s}}}
\newcommand{\aspot}[1]{\alpha_{\mathrm{s}}^{#1}}

\newcommand{\eT}{E_{\perp}}
\newcommand{\ki}[1]{k_{#1}}
\newcommand{\kT}{k_{\perp}}
\newcommand{\kTpot}[1]{k_{\perp}^{#1}}
\newcommand{\kTi}[1]{k_{{\perp},#1}}
\newcommand{\kTipot}[2]{k_{{\perp},#1} ^{#2}}
\newcommand{\qi}[1]{q_{#1}}
\newcommand{\qT}{q_{\perp}}
\newcommand{\qTpot}[1]{q_{\perp}^{#1}}
\newcommand{\qTi}[1]{q_{{\perp},#1}}
\newcommand{\shat}{\hat{s}}
\newcommand{\sigmahat}{\hat{\sigma}}
\renewcommand{\xi}[1]{x_{#1}}

\newlength{\abstwidth}
\begin{document}
 
\sloppy
 
\pagestyle{empty}
 
\begin{flushright}
LU TP 99--43 \\
December 1999
\end{flushright}
 
\vspace{\fill}
 
\begin{center}
{\LARGE\bf Minijets and Transverse Energy Flow}\\[3mm]
{\LARGE\bf in High Energy Collisions}\\[10mm]
{\Large G\"osta Gustafson\footnote{gosta@thep.lu.se}
and  Gabriela Miu\footnote{gabriela@thep.lu.se}}\\ [2mm]
{\it Department of Theoretical Physics,}\\[1mm]
{\it Lund University, Lund, Sweden}
\end{center}

\vspace{\fill}
\begin{center}
{\bf Abstract}\\[2ex]
\begin{minipage}{\abstwidth}
When studying the production of minijets and transverse energy flow
in high energy hadron-hadron or nucleus-nucleus collisions, two 
essential points have to be taken into account. First, one has to
account for the virtuality of the colliding partons and secondly, 
it is important to avoid double counting, when many links in a 
parton chain can be interpreted as the momentum transfer in a hard 
subcollision. The Linked Dipole Chain model, introduced for low-$x$
DIS, is particularly suitable for a study of these problems. It 
describes (mini)jet production in a $k_\perp$-factorizing 
formalism, which includes all links in a parton chain on an equal 
footing, avoiding double counting. In a ``naive'' calculation based
on integrated structure functions, the cross section blows up for 
small $p_\perp$, which makes it necessary to introduce a soft 
cutoff. In our approach we find a dynamical suppression at low 
$p_\perp$, which makes it possible to extrapolate to higher 
energies and make more reliable predictions for RHIC and LHC.

\end{minipage}
\end{center}

\vspace{\fill}
 
\clearpage
\pagestyle{plain}
\setcounter{page}{1}

\section{Introduction}
\label{sect-intro}

With increasing energies in hadron-hadron or nucleus-nucleus 
collisions, the cross section for hard subcollisions increases 
and becomes more and more important. The amount of minijets and 
transverse energy becomes essential for understanding the 
background in searches for new particles or new phenomena at 
the LHC as well as for the "initial conditions" in nucleus-nucleus
collisions. In calculations of the flow in either a quark-gluon 
plasma or a hadronic phase, the results are very sensitive to the
properties of the initial parton state which resulted from a large 
number of hard subcollisions. Thus a reliable estimate of these 
initial conditions is essential for the interpretation of signals
from a possible plasma formation in experiments at RHIC or LHC.
 
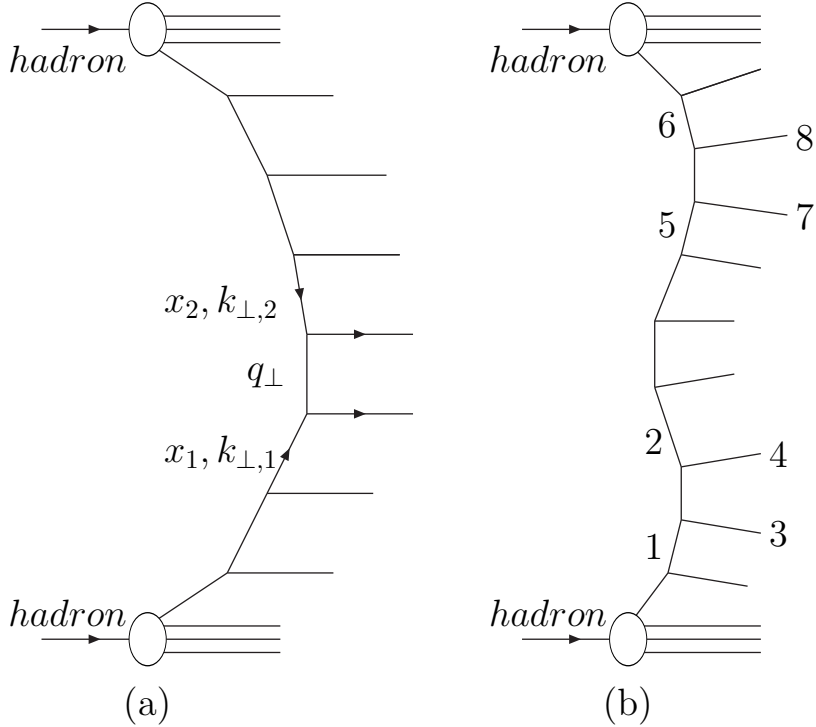
\begin{figure}[htb]
\begin{center}
\scalebox{1}{\mbox{
\begin{picture}(170,260)(0,0)
\ArrowLine(10,15)(50,15)
\Text(20,25)[]{\large $hadron$}
\Line(50,20)(100,20)
\Line(50,15)(100,15)
\Line(50,10)(100,10)
\Line(50,20)(80,40)
\GOval(50,15)(10,7)(0){1}
\Line(80,40)(120,40)
\Line(80,40)(95,70)
\Line(95,70)(135,70)
\ArrowLine(95,70)(110,100)
\ArrowLine(110,100)(150,100)
\Line(110,100)(110,130)
\ArrowLine(110,130)(150,130)
\Line(105,160)(145,160)
\ArrowLine(105,160)(110,130)
\Line(95,190)(105,160)
\Line(105,160)(145,160)
\Line(95,190)(140,190)
\Line(80,220)(95,190)
\Line(80,220)(120,220)
\Line(50,240)(80,220)
\ArrowLine(10,245)(50,245)
\Text(20,235)[]{\large $hadron$}
\Line(50,250)(100,250)
\Line(50,245)(100,245)
\Line(50,240)(100,240)
\GOval(50,245)(10,7)(0){1}
\Text(78,85)[]{\large $\xi{1},\kTi{1}$}
\Text(95,115)[]{\large $\qT$}
\Text(78,142)[]{\large $\xi{2},\kTi{2}$}
\Text(50,-10)[]{\large (a)}
\end{picture} }}
\scalebox{1}{\mbox{
\begin{picture}(170,260)(0,0)
\ArrowLine(10,15)(50,15)
\Text(20,25)[]{\large $hadron$}
\Line(50,20)(100,20)
\Line(50,15)(100,15)
\Line(50,10)(100,10)
\Line(50,20)(65,40)
\GOval(50,15)(10,7)(0){1}
\Line(65,40)(95,35)
\Line(65,40)(70,60)
\Line(70,60)(100,55)
\Line(70,60)(70,80)
\Line(70,80)(100,85)
\Line(70,80)(60,110)
\Line(60,110)(90,115)
\Line(60,110)(60,135)
\Line(60,135)(90,135)
\Line(60,135)(70,160)
\Line(70,160)(100,155)
\Line(70,160)(75,180)
\Line(75,180)(110,175)
\Line(75,180)(75,200)
\Line(75,200)(110,205)
\Line(75,200)(70,220)
\Line(70,220)(100,230)
\Line(70,220)(100,230)
\Text(60,50)[]{\large 1}
\Text(107,55)[]{\large 3}
\Text(60,90)[]{\large 2}
\Text(107,85)[]{\large 4}
\Text(65,172)[]{\large 5}
\Text(117,175)[]{\large 7}
\Text(117,205)[]{\large 8}
\Text(65,210)[]{\large 6}
\ArrowLine(10,245)(50,245)
\Text(20,235)[]{\large $hadron$}
\Line(50,250)(100,250)
\Line(50,245)(100,245)
\Line(50,240)(100,240)
\Line(70,220)(50,240)
\GOval(50,245)(10,7)(0){1}
\Text(50,-10)[]{\large (b)}
\end{picture} }}
\end{center}
 \caption{\label{fig:h-h}\em A fan diagram for a hadron-hadron 
collision.}
\end{figure}

At high $q_\perp$ the jet cross section can be described by a 
product of structure functions describing the flux of partons and 
the cross section for a hard partonic subcollision. Symbolically 
we write (cf. Fig.~\ref{fig:h-h}a)
\begin{equation}
\frac{d \sigma}{d q_\perp^2} \sim F(x_1,q_\perp^2)\,
F(x_2,q_\perp^2)\,\frac{d \hat{\sigma}}{d q_\perp^2}.
\label{naive}
\end{equation}
This is a relevant description when $q_\perp$ is so large that the 
structure functions can be described by DGLAP evolution, i. e. by 
$k_\perp$-ordered chains from the incoming hadrons towards the hard
subcollision. In the hard collision cross section, 
$d\hat{\sigma}/d q_\perp^2$, the colliding partons are then 
approximately on shell. For smaller $q_\perp^2$ and/or larger 
energies (which implies smaller $x$-values) we enter the BFKL 
region, in which non-ordered chains are important. This has two 
essential consequences:

i) The virtuality of one or both of the colliding partons 
($k_{\perp1}^2$ and/or $k_{\perp2}^2$ in Fig.~\ref{fig:h-h}a) can 
be larger than the momentum transfer $q_\perp^2$. This implies that
the virtualities of the colliding partons have to be taken into 
account. This can be achieved by means of nonintegrated structure 
functions and off-shell subcollision cross sections.

ii) In a chain with non-ordered $k_\perp$-values a single hard 
subcollision cannot be isolated. There may be several links in a 
single chain, which can be regarded as a hard collision. This 
situation can be analyzed in a formalism in which every link in 
Fig.~\ref{fig:h-h} can be regarded as a hard subcollision. This 
implies that each final state parton is active in two different 
subcollisions. (The number of links in the chain determines both 
the number of subcollisions and the number of final state partons.)
Therefore special care has to be taken to avoid double counting.
%

A formalism for DIS which interpolates between the DGLAP and BFKL 
regions, and which describes both the cross section and the final 
state properties, was developed by Ciafaloni, Catani, Fiorani, 
and Marchesini \cite{CCFM}. This or equivalent formalisms can be 
used e. g. to calculate the production of heavy quarks in hadronic 
collisions. In this reaction there is a single link which contains 
the heavy quark, and the CCFM formalism or the semihard approach 
can be used to calculate the unordered evolution towards the heavy 
quark from the projectile and the target ends \cite{referee}. Thus 
these formalisms can be used to solve problem i) above.

For non-$k_\perp$-ordered chains the single diagram in 
Fig.~\ref{fig:h-h} can correspond to many hard subcollisions, 
e. g. $1+2\rightarrow3+4$ and $5+6\rightarrow7+8$ in 
Fig.~\ref{fig:h-h}b. For the first reaction, $1+2\rightarrow3+4$, 
the section of the ladder between partons 2 and 5 should be 
regarded as an evolution from the top (projectile) end, while for 
the reaction $5+6\rightarrow7+8$ the same section should be 
regarded as an evolution from the lower (target) end of the chain. 
A description of this process is most simple in a formalism in 
which the evolution is explicitly symmetric between the two ends. 
It is also essential that the formalism can be interpreted in 
terms of production probabilities for exclusive final states. 
The symmetry is not a trivial feature for the following reason. 
The fan diagrams in Fig.~\ref{fig:h-h} should be regarded as the 
initial state radiation. To get the complete final state, final 
state emission has to be added in specified kinematic regions. 
The separation between initial state and final state emission is 
not determined by Nature, but is defined by the calculation scheme,
and therefore the regions for final state emission depend on the 
formalism used. In the CCFM approach this separation is not 
symmetric, and consequently also the initial state ladder is 
asymmetric. Thus the CCFM formalism is not immediately convenient 
to describe many subcollisions in a single chain, and thus to solve
problem ii) above.

The final state emission, when one gluon is split into two, should 
not significantly affect the flow of energy. Note, however, that 
the distribution of produced gluons is not a uniquely defined 
quantity. Besides the arbitrariness in the separation between 
initial and final state emission, discussed above, also the final 
state radiation depends on the resolution chosen in the definition 
of an individual gluon, i. e. on the cut-off in the final state 
cascades.

The Linked Dipole Chain (LDC) model \cite{LDC,Hamid} is a 
reformulation and generalization of the CCFM formalism. The 
separation between initial and final state radiation is chosen in 
such a way that the description is explicitly symmetric between 
the projectile and target ends of the ladder. It can within a 
single formalism describe different types of events, "normal DIS", 
boson-gluon fusion events, and events with hard subcollisions in 
resolved photon events. For such a resolved photon event the result
can be directly interpreted as evolutions from both the proton and 
the photon ends towards a hard parton-parton collision. This 
feature implies that the LDC model is also particularly suitable 
for a description of hard collisions in hadronic interactions. For 
any link in the ladder, the contribution can be directly 
interpreted as evolution from the two ends multiplied by an 
off-shell parton-parton cross section for the link under study. 
This symmetry property of the LDC model makes it useful in solving 
problem ii) above (the double-counting is easily corrected for in 
this formalism).

It is well known that the BFKL formalism with a constant coupling, 
$\alpha_s$, implies that the transverse momenta grow like a random 
walk in $y$ or in $\ln 1/x$ (what is called Bartels' cigar). For a 
running $\alpha_s$ we obtain instead a saturation of the 
$k_\perp$-distribution \cite{Hamid}. Therefore very large chains 
develop a central plateau in rapidity. This means that with 
increasing beam energy the transverse energy density in each chain 
will stay limited, although the total $dE_\perp/dy$ will increase 
as the number of possible chains grows with energy. The fact that 
the $k_\perp$-distribution stays limited also for high energies 
implies that the result is sensitive to soft nonperturbative 
physics, and thus depends on a necessary cut-off for small 
$k_\perp$. This cut-off can, however, be fixed by a 
phenomenological fit to experimental data for $F_2(x,Q^2)$, 
and therefore does not imply a large extra uncertainty.

When comparing our result with the ``naive'' expression in 
eq.~(\ref{naive}) we find that the latter significantly 
overestimates the cross section for smaller $q_\perp$. In order 
to agree with experimental data for $E_\perp$-flow, many 
calculations based on eq.~(\ref{naive}) introduce a cutoff for 
small $q_\perp$ of the order 2 GeV/c \cite{cutoff}. This implies, 
however, that it is difficult to make predictions for higher 
energies. Without a dynamical understanding of the origin of the 
overestimate, it is not possible to judge how an effective cut-off 
should vary with energy. The dynamical suppression in our formalism
corresponds to such an effective cut-off, indeed of the order 
2 GeV/c, which is growing slowly with the total collision energy.

The results in this paper are mainly based on approximate analytic 
calculations, which demonstrate the qualitative features. In a 
future publication we want to present more detailed results based 
on MC simulations, where e. g. corrections from quark lines, 
subleading terms in the splitting functions and exact 
energy-momentum conservation, are also taken into account.

\section{DIS}
\label{sect-DIS}

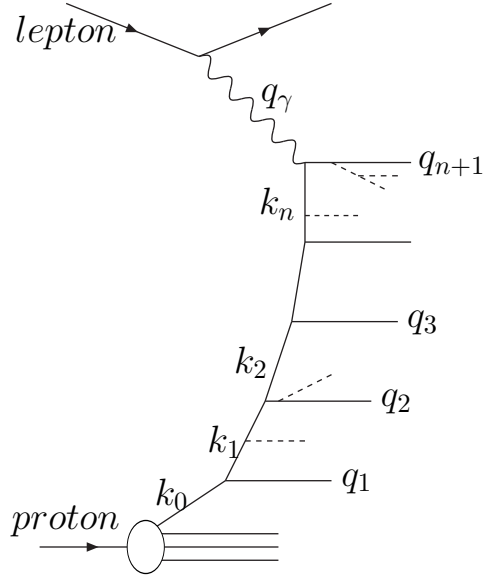
\begin{figure}[htb]
\begin{center}
\scalebox{1}{\mbox{
\begin{picture}(170,230)(0,0)
\ArrowLine(10,15)(50,15)
\Text(20,25)[]{\large $proton$}
\Line(50,20)(100,20)
\Line(50,15)(100,15)
\Line(50,10)(100,10)
\Line(50,20)(80,40)
\GOval(50,15)(10,7)(0){1}
\Text(60,35)[]{\large $k_{0}$}
\Text(130,40)[]{\large $q_{1}$}
\Line(80,40)(120,40)
\Line(80,40)(95,70)
\Text(80,55)[]{\large $k_{1}$}
\Text(145,70)[]{\large $q_{2}$}
\DashLine(87.5,55)(110,55){2}
\Text(90,85)[]{\large $k_{2}$}
\Text(155,100)[]{\large $q_{3}$}
\Line(95,70)(135,70)
\Line(95,70)(105,100)
\DashLine(100,70)(120,80){2}
\Line(105,100)(145,100)
\Line(105,100)(110,130)
\Line(110,130)(150,130)
\Line(110,130)(110,160)
\DashLine(110,140)(130,140){2}
\Line(110,160)(150,160)
\DashLine(120,160)(140,150){2}
\DashLine(130,155)(145,155){2}
\ArrowLine(20,220)(70,200)
\Text(166,160)[]{\large $q_{n+1}$}
\Text(100,145)[]{\large $k_{n}$}
\Text(20,210)[]{\large $lepton$}
\Text(100,185)[]{\large $q_{\gamma}$}
\ArrowLine(70,200)(120,220)
\Photon(70,200)(110,160){3}{5}
\end{picture} }}
\end{center}
 \caption{\label{fig:fanf}\em A fan diagram for a DIS event. 
The quasireal partons from the initial state radiation are 
denoted $q_i$, and the virtual propagators $k_i$. The dashed 
lines denote final state radiation.}
\end{figure}

A deep inelastic scattering event is generally described 
in terms of a fan diagram as shown in Fig~\ref{fig:fanf}. 
Here $q_i$ denote quasireal partons emitted as initial state 
radiation, while the links $k_i$ are virtual. The dashed lines 
denote final state emission, which is assumed to be emitted without
changing the cross section and with negligible recoils 
for the emitting partons $q_i$.
%

In the large $Q^2$ region, the DGLAP region, the dominant 
contributions come from ordered chains which satisfy
$
Q^2 >k_{\perp,n}^2 > ...> k_{\perp,i}^2 > k_{\perp,i-1}^2$ and $
k_{+,i} > k_{+,i+1}
$.
Each such chain gives a contribution 
($x_i \equiv k_{+,i}/ P_{+,tot}$)
\begin{equation}
\prod_i^n \bar{\alpha} \frac{dx_i}{x_i} 
\frac{dk_{\perp,i}^2}{k_{\perp,i}^2} \,\,\,\,
{\mathrm {where}} \,\,\,\, \bar{\alpha} \equiv \frac{3\alpha_s}
{\pi}.
\end{equation}
Integrating over the appropriate integration regions and summing 
over possible values of $n$, the number of links in the chain, 
we readily obtain (for a fixed coupling $\bar{\alpha}$)
\begin{equation}
 F(x,Q^2) \sim \sum_n \bar{\alpha}^n \frac{(\ln1/x)^n}{n!} 
\frac{(\ln Q^2)^n}{n!} 
 \approx \exp (2\sqrt{\bar{\alpha} \ln Q^2 \ln 1/x}).
\label{fDGLAP}
\end{equation}
For a running coupling we get instead of $\ln Q^2$ a factor 
$\ln \ln Q^2$. We want to stress that to get the properties of the 
{\em final state}, we have to add final state emission within 
regions allowed by angular ordering.

For very small $x$ and limited $Q^2$, the BFKL region, also 
non-ordered chains become important, although suppressed. 
Solutions to the BFKL equation \cite{BFKL} increase like a power 
$1/x^\lambda$ for small $x$-values. Such a power-like behavior 
with $\lambda \sim 0.3$, is indicated in data on $F_2$ from HERA. 
(We note, however, that it is also possible to describe this 
increase by NLO DGLAP evolution.)

In the interpolation region between the DGLAP and BFKL regimes we 
have to calculate suppressed contributions from non-ordered chains.
For each chain of initial state radiation, final state emission 
should be added within specified kinematic regions. This final 
state radiation should give negligible recoils. It should also be 
described by Sudakov form factors, which implies that the final 
state emission only affects the properties of the final state and 
not the cross section (i. e. the reaction probability), 
which is described by the structure function. Thus a chain with 
specified initial state radiation represents a set of final states 
with all possible final state emissions within the allowed regions.
We want to stress that the separation between initial and final 
state radiation is not given by Nature, but is defined by the 
calculation scheme. 

A specific scheme, which interpolates between the DGLAP and the 
BFKL results, was presented by Ciafaloni, Catani, Fiorani and 
Marchesini, the CCFM model \cite{CCFM}. In this scheme those final 
state gluons, which are not followed in rapidity (or angle) by a 
more energetic gluon, are regarded as initial state radiation, all 
other as final state emission. With this definition they showed 
that for small $x$ the nonintegrated structure function, $f$, is 
given by the expression (see ref \cite{CCFM} for details)
\begin{equation}
 f(x,k_\perp^2,q^2) \sim \sum_n \int \prod^n \bar{\alpha} 
\frac{dz_i}{z_i} \frac{d^2 q_{\perp,i}}{\pi q_{\perp,i}^2} 
\Delta_{ne} (z_i, k_{\perp,i}^2, q_i^2) \delta(x-\Pi z_i) 
\delta(k_\perp^2-k_{\perp,n}^2) \delta(q^2-q_n^2).
\label{fCCFM}
\end{equation}
Here $q_{\perp,i}$ and $k_{\perp,i}$ are the transverse momentum of
the real and virtual partons as indicated in Fig.~\ref{fig:fanf}, 
and the variable $q$ is defined by the relation 
$q \equiv q_\perp/(1-z)$, and the function 
$\Delta_{ne} (z, k_{\perp}^2, q^2)$ is a specific non-eikonal form 
factor. The integration region is restricted by the relation
\begin{equation}
 q_{\perp,i+1} > z_i q_{\perp,i} / (1-z_i)
\label{ao}
\end{equation}
which follows from angular ordering, and by the kinematic 
constraint (called the consistency constraint)
\begin{equation}
 k_{\perp,i}^2 > z_i q_{\perp,i}^2.
\label{cc}
\end{equation}
We note that this nonintegrated structure function
$f(x,k_\perp^2,q^2)$ depends on two scales, $k_\perp$ which defines
the transverse momentum of the last link in the chain, and $q$ 
which depends on the angle of the last emission, and therefore 
specifies the boundary of the angular region in which later 
emissions are allowed.

\section{The Linked Dipole Chain Model}
\label{sect-LDC}

The Linked Dipole Chain model \cite{LDC} (LDC) is a reformulation 
and generalization of the CCFM model. In the LDC model more gluons 
are treated as final state radiation. The remaining (initial state)
gluons are ordered both in $q_+$ and in $q_-$. (This implies that 
they are also ordered in angle or rapidity $y$.) 
Thus a single chain in the LDC model represents a set of chains in 
the CCFM scheme, all with the same ``backbone'' of harder gluons. 
In ref \cite{LDC} it is demonstrated that if we sum over all states
in this set, with their corresponding non-eikonal form factors, 
then all this adds up to unity. Thus the form factors are exactly 
canceled, and a nonintegrated structure function ${\cal F}$ can be 
written in the simple form
\begin{equation}
 {\cal F}(x,k_\perp^2) \sim \sum_n \int \prod^n \bar{\alpha} 
\frac{dz_i}{z_i} \frac{d^2 q_{\perp,i}}{\pi q_{\perp,i}^2} 
\theta(q_{+,i-1} -q_{+,i}) \theta(q_{-,i} -q_{-,i-1}) 
\delta(x-\Pi z_i) \delta(\ln k_\perp^2 - \ln k_{\perp,n}^2).
\label{fq}
\end{equation}
We note in particular that this result is symmetric in $q_+$ and 
$q_-$. (An essential point for this symmetry and the cancelation 
of the non-eikonal form factors is the consistency constraint in 
eq.~(\ref{cc}).) The fact that the ordering in both $q_+$ and $q_-$
also automatically implies an ordering in $y$ (i. e. in angle), 
means that the nonintegrated structure function 
${\cal F}(x,k_\perp^2)$ depends only on {\em one} scale, 
$k_\perp^2$. (In contrast, the corresponding form factor, 
$f(x,k_\perp^2,q^2)$, in the CCFM model also depends on the scale 
$q^2$, related to the boundary of the angular region allowed for 
further emissions.) This fact implies a considerable simplification
of the formalism. The symmetry also implies that the chain in 
Fig~\ref{fig:fanf} can be interpreted either as evolving from 
bottom to top (i. e. from the proton end towards the photon) or 
evolving from top to bottom (from the photon towards the proton). 
This feature will be essential for our analysis of the minijet 
distribution.

%
\begin{figure}
\begin{center}
%
\scalebox{0.9} {\mbox{
\begin{picture}(140,230)(0,-50)
\Line(10,15)(50,15)
\Line(50,20)(80,20)
\Line(50,15)(80,15)
\Line(50,10)(80,10)
\Line(50,15)(60,40)
\GOval(50,15)(10,7)(0){1}
\Line(60,40)(90,40)
\Line(60,40)(70,70)
\Line(70,70)(105,70)
\Line(70,70)(75,100)
\Line(75,100)(110,100)
\Line(75,100)(80,130)
\Line(80,130)(110,130)
\Line(80,130)(85,160)
\Line(85,160)(120,160)
\Line(85,160)(85,190)
\Line(85,190)(120,190)
\Photon(50,210)(85,190){3}{4}
\Line(85,190)(120,190)
\Line(20,220)(50,210)
\Line(50,210)(80,220)
\Text(47,34)[]{\large $k_{0}$}
\Text(55,55)[]{\large $k_{1}$}
\Text(65,85)[]{\large $k_{2}$}
\Text(70,115)[]{\large $k_{3}$}
\Text(75,145)[]{\large $k_{4}$}
\Text(75,173)[]{\large $k_{5}$}
\Text(100,40)[]{\large $q_{1}$}
\Text(115,70)[]{\large $q_{2}$}
\Text(120,100)[]{\large $q_{3}$}
\Text(125,130)[]{\large $q_{4}$}
\Text(130,160)[]{\large $q_{5}$}
\Text(130,190)[]{\large $q_{6}$}
\end{picture}}}
%
\scalebox{0.9} {\mbox{
\begin{picture}(340,280)(0,0)
\Line(40,20)(300,20)
\Line(40,20)(170,280)
\Line(170,280)(300,20)
\Line(80,100)(120,20)
\Text(60,85)[]{$q_{6}$}
\Vertex(70,80){2}
\Text(110,70)[]{$k_{5}$}
\Vertex(125,80){2}
\Text(125,90)[]{$q_{5}$}
\Line(90,80)(125,80)
\Line(125,80)(130,70)
\Line(130,70)(145,70)
\Text(140,60)[]{$k_{4}$}
\Line(145,70)(160,100)
\Line(160,100)(190,100)
\Vertex(160,100){2}
\Text(160,110)[]{$q_{4}$}
\Text(175,90)[]{$k_{3}$}
\Vertex(190,100){2}
\Text(190,110)[]{$q_{3}$}
\Line(190,100)(200,80)
\Line(200,80)(230,80)
\Text(215,70)[]{$k_{2}$}
\Line(230,80)(250,40)
\Vertex(230,80){2}
\Text(230,90)[]{$q_{2}$}
\Line(250,40)(290,40)
\Text(270,30)[]{$k_{1}$}
\Vertex(290,40){2}
\Text(300,50)[]{$q_{1}$}
\LongArrow(250,180)(250,230)
\LongArrow(250,180)(300,180)
\Text(250,240)[]{$\ln \kTpot{2}$}
\Text(310,180)[]{$y$}
\DashLine(125,80)(125,20){2}
\Line(125,80)(135,10)
\Line(125,20)(135,10)
\DashLine(160,100)(160,20){2}
\Line(160,100)(180,0)
\Line(180,0)(160,20)
\DashLine(190,100)(190,20){2}
\Line(190,100)(210,0)
\Line(210,0)(190,20)
\DashLine(230,80)(230,20){2}
\Line(230,80)(240,10)
\Line(240,10)(230,20)
\DashLine(40,-10)(40,20){2}
\DashLine(120,-10)(120,20){2}
\DashLine(300,-10)(300,20){2}
\LongArrow(40,-5)(120,-5)
\LongArrow(120,-5)(40,-5)
\LongArrow(120,-5)(300,-5)
\LongArrow(300,-5)(120,-5)
\Text(80,-15)[]{$\ln Q^2$}
\Text(210,-15)[]{$\ln 1/x$}
\LongArrow(220,130)(260,150)
\Text(280,150)[]{$\ln q_+$}
\LongArrow(120,130)(80,150)
\Text(60,150)[]{$\ln q_-$}
\DashLine(15,100)(80,100){2}
\DashLine(15,20)(40,20){2}
\LongArrow(20,20)(20,100)
\LongArrow(20,100)(20,20)
\Text(0,60)[]{$\ln Q^2$}
\end{picture}}}
\end{center}
 \caption{\label{fig:typ-chain}\em The initial state emissions 
$q_i$ in the $(y,\kappa=ln(k_{\perp}^2))$-plane. Final state 
radiation is allowed in the region below the horizontal lines.}
\end{figure}
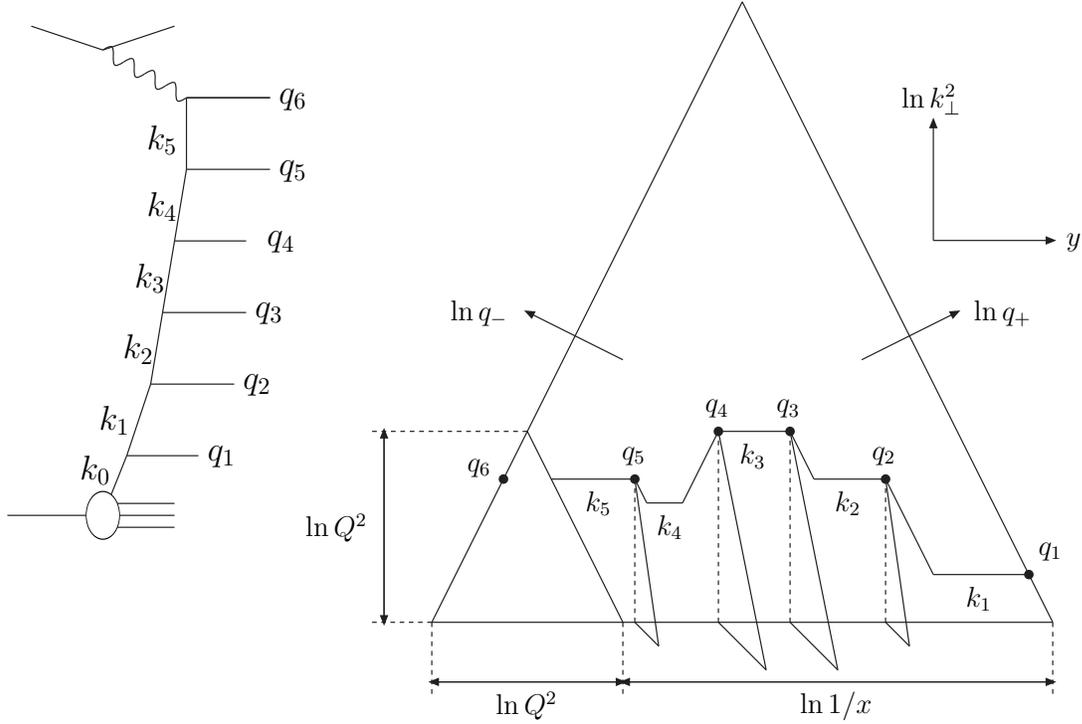

Fig.~\ref{fig:typ-chain} shows a typical chain in a 
$(y,\ln q_{\perp}^2)$-plane. The real emitted partons $q_i$ 
are mapped onto points in this figure. The lightcone components 
$\ln q_\pm = \pm y + \frac{1}{2} \ln q_\perp^2$ grow in the upper 
right and upper left directions as indicated in the figure. The 
virtual propagators do not have a well defined rapidity; they are 
represented by horizontal lines, whose left ends correspond to the 
values of $k_{\perp,i}$ and $k_{+,i}$, while the right ends 
correspond to $k_{\perp,i}$ and $k_{-,i}$. 
The larger region allowed for final state radiation in the LDC 
model corresponds to the region below the horizontal lines in 
Fig.~\ref{fig:typ-chain}.

It is convenient to express the result 
in terms of the propagator momenta $k_i$, using the relations 
$d^2 q_{\perp,i} = d^2 k_{\perp,i}$ and 
$q_{\perp,i}^2 \approx \max( k_{\perp,i}^2, k_{\perp,i-1}^2)$.
Thus eq.~(\ref{fq}) can be written in the form
\begin{equation}
 {\cal F} \sim \sum \int \prod \bar{\alpha} \frac{dz_i}{z_i} 
\frac{d k_{\perp,i}^2}{ \max( k_{\perp,i}^2, k_{\perp,i-1}^2)}.
\label{fk}
\end{equation}
where we have suppressed the $\theta$- and $\delta$-functions. 
This result implies that for a ``step up'' or a ``step down'' in 
$k_{\perp}$ we find the following weights
\begin{eqnarray} 
\frac{d^2 q_{\perp,i}}{q_{\perp,i}^2} \approx \frac{d^2 k_{\perp,i}}
{k_{\perp,i}^2},
\,\,\,\, k_{\perp,i} > k_{\perp,i-1} \,\,\,\,\,{\mathrm {and}}
\label{step-up}\\
\frac{d^2 q_{\perp,i}}{q_{\perp,i}^2} \approx \frac{d^2 k_{\perp,i}}
{k_{\perp,i}^2} 
\cdot \frac{ k_{\perp,i}^2}{k_{\perp,i-1}^2},\,\,\,\,\,
k_{\perp,i} < k_{\perp,i-1}.
\label{step-down}
\end{eqnarray}
Thus for a step down we have an extra suppression factor 
$ k_{\perp,i}^2/k_{\perp,i-1}^2$. This implies that if the chain 
goes up to $k_{\perp,max}$ and then down to $k_{\perp,final}$ we 
obtain the factor
\begin{equation}\prod  \frac{d k_{\perp,i}^2}{k_{\perp,i}^2} \cdot 
\frac{ k_{\perp,final}^2}{k_{\perp,max}^2}.
\end{equation}
This gives a factor $1/k_{\perp,max}^4$, which corresponds to a 
hard parton-parton subcollision. 

\begin{figure}[t]
\begin{center}
\mbox{\epsfig{file=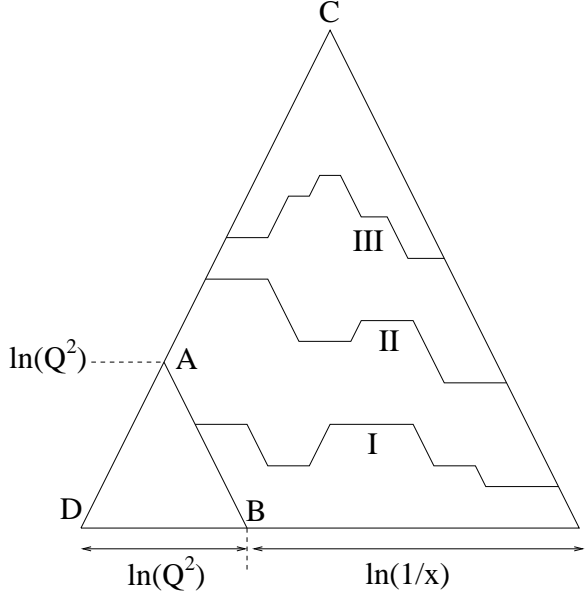,width=7.7cm}}
\end{center}
\caption{\label{fig:diff-proc}\em Different types of parton chains. 
I: ``Normal'' DIS. II: Boson-gluon fusion. III: Hard resolved 
photon collision.}
\end{figure}

In DIS also chains for which $k_{\perp,final}^2 > Q^2$ can 
contribute to the cross section. These chains correspond to 
boson-gluon fusion events, and the definition of the structure 
function $F(x,Q^2)$ implies that these contributions contain a 
similar suppression factor $Q^2/k_{\perp,final}^2$.
Thus the relation between the integrated and the nonintegrated 
structure functions, sometimes symbolically written as 
$F(x,Q^2) \sim \int \frac{d k_{\perp}^2}{k_{\perp}^2} 
{\cal F}(x,k_{\perp}^2)$, is more explicitly given by the relation
\begin{equation}
F(x,Q^2) = \int^{Q^2} \frac{d k_{\perp}^2}{k_{\perp}^2} {
\cal F}(x,k_{\perp}^2) + \int_{Q^2} \frac{d k_{\perp}^2}
{k_{\perp}^2} {\cal F}(x \frac{k_\perp^2}{Q^2},k_{\perp}^2) 
\frac{Q^2}{k_{\perp}^2}.
\label{relation-new}
\end{equation}
In the second term, besides the suppression factor 
$Q^2 / k_{\perp}^2$ we have also a shifted $x$-value. `
`Normal'' chains, for which $k_{\perp,final}^2 < Q^2$, end on the 
line AB in Fig.~\ref{fig:diff-proc}, which corresponds to 
$k_{+,final} = x \cdot P_{+,tot}$, but due to energy-momentum 
conservation, chains for which $k_{\perp,final}^2 > Q^2$ end 
instead on the line AC, which corresponds to 
$k_{+,final} = x \cdot \frac{k_{\perp,final}^2}{Q^2} \cdot 
P_{+,tot}$. Note that the relation in eq.~(\ref{relation-new}) 
is a leading approximation valid when $k_\perp^2$ is either much 
smaller or much larger than $Q^2$. Exact energy-momentum 
conservation implies that the argument in ${\cal F}$ should be 
$x(k_\perp^2+Q^2)/Q^2$, which approaches $x$ or $x k_\perp^2 / Q^2$
in the two limits. The exact kinematic relation is included in the 
MC in the whole chain including this final vertex. Our experience 
indicates that the approximations presented here are sufficient to 
understand the essential properties.

We see that in the LDC model different types of reactions are 
treated in the same formalism, without double counting or missed 
parts of phase space. Thus Fig.~\ref{fig:diff-proc} shows three 
chains representing different types of reactions:

  I. ``normal'' DIS,

  II. boson-gluon fusion: $k_{\perp,final}^2 > Q^2$,

  III. hard resolved photon collision: 
     $k_{\perp,max}^2 > k_{\perp,final}^2 (> Q^2)$.

As mentioned above, the LDC formalism is fully 
{\em left-right symmetric}, meaning that the same result is 
obtained if the chain is generated from the photon end instead of 
from the proton end. Although not evident from eq.~(\ref{fk}), 
this is obvious from the expression in eq.~(\ref{fq}). This feature
also means that the same formalism can be used if the (resolved) 
photon in one end of the chain is replaced by a hadron. 
In section 5 we will see that the symmetric property of the LDC 
formalism makes it particularly effective for an 
understanding of the minijet distribution in hadronic reactions or 
nucleus-nucleus collisions. It is also important that the result in
eq.~(\ref{fq}) not only describes inclusive properties, but can be 
interpreted as the production probability for an exclusive final 
state.

Before we go into more details about the properties of this minijet
plateau, we want to mention some other properties of the LDC model:
\begin{itemize}
\item It is straight forward to allow for a running coupling 
$\alpha_s$ within the formalism.
\item It is possible to include quarks and other non-leading 
effects (non-leading terms in the splitting functions and exact 
energy-momentum conservation). The result is also improved when the
link with highest $p_\perp$ is adjusted to the exact matrix element.
\item The chains contain only those gluons (or quarks), which cannot
be regarded as final state radiation. This might be called the 
``backbone'' of the chain, and the fact that hard jets are not 
yet subdivided into many sub-jets may make it more easy to study
minijets and $E_\perp$-flow, and to interpret the results of the 
calculations.
\item The fact that there are fewer gluons in the primary chain 
also implies that typical $z$-values are smaller, and therefore 
smaller sub-leading  effects are expected.
\item The formalism is suitable for MC simulation. Such a program 
is developed by L\"onnblad and Kharraziha \cite{LDCMC}.
\end{itemize}
Naturally it is essential to verify that the LDC model also can 
reproduce experimental data. Preliminary results indicate that the 
LDC MC indeed is able to successfully describe experimental 
results, both for the structure functions and for the properties 
of the final states, for example the production of jets and 
transverse energy flow \cite{prel}.

\section{Integral equations and asymptotic behavior}
\label{sect-integral}

It is straight forward to derive integral equations for the 
non-integrated structure function ${\cal F}$. From the relation in 
eq.~(\ref{relation-new}) we obtain the equation
\begin{equation}
\frac{\partial {\cal F}(l,\kappa)}{\partial l} = 
\int_{\kappa_0}^\kappa d\kappa' 
\bar{\alpha}(\kappa) {\cal F}(l,\kappa') \nonumber \\+ 
\int_\kappa d\kappa' 
\bar{\alpha}(\kappa') {\cal F}(l+\kappa-\kappa',\kappa') 
exp[-(\kappa' - \kappa)]
\label{diff-eq}
\end{equation}
where $l \equiv \ln (1/x)$ and $\kappa \equiv \ln k_{\perp}^2$.

It is well known that the BFKL formalism with a constant coupling, 
$\alpha_s$, implies that the transverse momenta grow like a random 
walk in $\ln k_\perp^2$, which gives a Gaussian distribution that 
widens with $\ln(1/x)$, 
$<\!\!\ln k_{\perp}^2\!\!>\, \sim \sqrt{\ln(1/x)}$.
A running coupling  favors smaller $k_{\perp}$-values, and this 
implies that the $k_{\perp}$-distribution does not widen 
indefinitely, but saturates for small $x$ \cite{Hamid}. 
The solution to eq.~(\ref{diff-eq}) can then be written in the 
factorized form
\begin{equation}
{\cal F} \approx \frac{1}{x^\lambda} \cdot f(\kappa), 
\,\,\,\,\,x \,\, {\mathrm {small}}.
\label{fact}
\end{equation}
The fact that small $k_\perp$-values are not suppressed implies that
the evolution and the value of $\lambda$ is sensitive to the soft 
region, and therefore cannot be determined from perturbative QCD 
alone. (This feature is consistent with the very large lower order 
corrections to the BFKL equation \cite{2ordBFKL}, which indicates 
that the perturbative series may converge badly.) This implies that
some kind of cutoff, $Q_0$, has to be introduced, and determined 
from fits to experimental data. If this cutoff is determined from 
fits to data on $F_2$ from DIS, then this result can be used in our
calculation of the minijet distribution in hadronic collisions. 
Although the dependence on $x$ and $k_\perp$ factorizes as 
expressed in eq.~(\ref{fact}), the two factors are not independent.
The relation in eq.~(\ref{diff-eq}) implies a correlation such that
the power $\lambda$ also determines the $k_\perp$-dependence. Thus 
for large $k_\perp$, $f$ takes the asymptotic form
\begin{equation}
f(\kappa) \propto \kappa^{\frac{\alpha_0}{\lambda} - 1}
\label{asymptotic}
\end{equation}
where $\alpha_0$ is defined by the relation
\begin{equation}
\bar{\alpha} \equiv \frac{3\alpha_s}{\pi} \equiv \frac{\alpha_0}
{\ln(Q^2/\Lambda^2)} \,\,\, \Rightarrow \,\,\,\alpha_0 = \frac{36}
{33-2n_f}. 
\label{alfa0}
\end{equation}
Actually MC calculations show that the simple power in 
eq.~(\ref{asymptotic}) is a surprisingly good approximation, not 
only for large values of $k_\perp$ (or 
$\kappa=\ln(k_\perp^2/\Lambda^2)$) but for the whole 
$k_\perp$-interval \cite{Hamid}.
 
The factorized form in eq.~(\ref{fact}) implies also, that for 
limited $Q^2$ and very small $x$ (or hadronic collisions at very 
high energies) a central plateau is developed in the minijet 
distribution. The properties of this plateau will be further 
studied in the next section.

\section{Inclusive Jet Cross Section}
\label{sect-incl}

In this section we present asymptotic results and general 
properties. A more detailed analysis will be presented in a future 
publication.

The properties of the fan diagram in Fig~\ref{fig:fanf} depend on 
the chosen separation between initial state and final state 
radiation. As mentioned above, the same event corresponds to a 
chain with fewer links in the LDC formalism than in e. g. the CCFM 
formalism, because in the LDC model more emissions are formally 
treated as final state radiation. The number of jets is not a 
well-defined concept, if it is not accompanied by some 
specification of the resolution, and two different schemes can be 
equally correct if one jet in one scheme corresponds to two or more
sub-jets in the other. The total $E_\perp$-flow is however expected
to be approximately unaffected, when one jet is split in two or 
several smaller jets by final state radiation. As mentioned in
section 3, the LDC formalism is fully symmetric with respect to 
the two ends of the fan or ladder diagram, and we will see that 
this makes it particularly effective for an analysis of the minijet
distribution. (In the CCFM formalism the fan diagram for the 
initial state radiation is not symmetric; the symmetry is restored 
only after inclusion of the final state emission.) It is also 
important that the formalism can be interpreted in terms of 
production probabilities for exclusive final states.

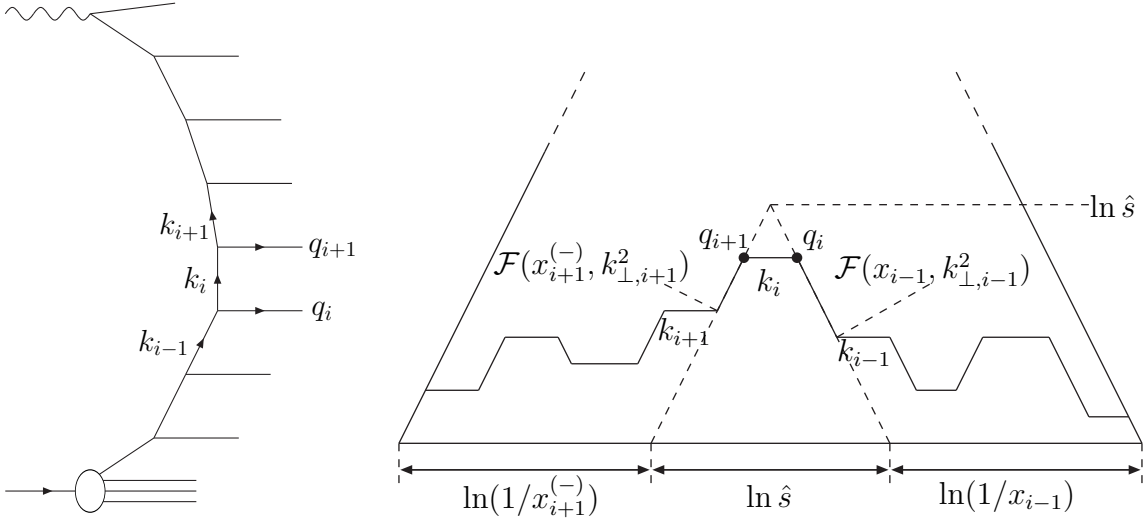
\begin{figure}[t]
\begin{center}
%
\scalebox{0.8}{\mbox{
\begin{picture}(170,260)(0,0)
\ArrowLine(10,15)(50,15)
\Line(50,20)(100,20)
\Line(50,15)(100,15)
\Line(50,10)(100,10)
\Line(50,20)(80,40)
\GOval(50,15)(10,7)(0){1}
\Line(80,40)(120,40)
\Line(80,40)(95,70)
\Line(95,70)(135,70)
\ArrowLine(95,70)(110,100)
\ArrowLine(110,100)(150,100)
\ArrowLine(110,100)(110,130)
\ArrowLine(110,130)(150,130)
\Line(105,160)(145,160)
\ArrowLine(110,130)(105,160)
\Line(95,190)(105,160)
\Line(105,160)(145,160)
\Line(95,190)(140,190)
\Line(80,220)(95,190)
\Line(80,220)(120,220)
\Line(50,240)(80,220)
\Photon(10,240)(50,240){3}{3}
\Line(50,240)(90,245)
\Text(85,85)[]{\large $\ki{i-1}$}
\Text(100,115)[]{\large $\ki{i}$}
\Text(95,140)[]{\large $\ki{i+1}$}
\Text(165,130)[]{\large $\qi{i+1}$}
\Text(160,100)[]{\large $\qi{i}$}
\end{picture} }}
\scalebox{1.}{\mbox{
\begin{picture}(300,170)(0,0)
\Line(10,30)(65,140)
\DashLine(65,140)(80,170){4}
\DashLine(220,170)(235,140){4}
\Line(235,140)(290,30)
\Line(10,30)(290,30)
\Line(20,50)(40,50)
\Line(40,50)(50,70)
\Line(50,70)(70,70)
\Line(70,70)(75,60)
\Line(75,60)(100,60)
\Line(100,60)(110,80)
\Line(110,80)(130,80)
\Line(130,80)(140,100)
\Line(140,100)(160,100)
\Line(160,100)(175,70)
\Line(175,70)(195,70)
\Line(195,70)(205,50)
\Line(205,50)(220,50)
\Line(220,50)(230,70)
\Line(230,70)(255,70)
\Line(255,70)(270,40)
\Line(270,40)(285,40)
\Vertex(140,100){2}
\Vertex(160,100){2}
\Text(133,107)[]{$\qi{i+1}$}
\Text(167,107)[]{$\qi{i}$}
\DashLine(105,30)(150,120){3}
\DashLine(195,30)(150,120){3}
\DashLine(150,120)(270,120){3}
\Text(280,120)[]{$\ln \shat$}
\DashLine(10,15)(10,30){3}
\DashLine(105,15)(105,30){3}
\DashLine(195,15)(195,30){3}
\DashLine(290,15)(290,30){3}
\LongArrow(12,20)(103,20)
\LongArrow(103,20)(12,20)
\LongArrow(107,20)(193,20)
\LongArrow(193,20)(107,20)
\LongArrow(197,20)(288,20)
\LongArrow(288,20)(197,20)
\Text(118,72)[]{$\ki{i+1}$}
\Text(186,64)[]{$\ki{i-1}$}
\Text(150,91)[]{$\ki{i}$}
\Text(60,10)[]{$\ln(1/x_{i+1}^{(-)})$}
\Text(150,10)[]{$\ln\shat$}
\Text(240,10)[]{$\ln(1/x_{i-1})$}
\Text(83,98)[]{$\mathcal{F}(x_{i+1}^{(-)},k_{{\perp},i+1} ^{2} )$}
\DashLine(110,90)(130,80){3}
\Text(212,95)[]{$\mathcal{F}(x_{i-1},k_{{\perp},i-1} ^{2} )$}
\DashLine(175,70)(210,90){3}
\end{picture} }}
\end{center}
\caption{\label{fig:long-chain}\em A local maximum transverse 
momentum, $k_{\perp,max} = k_{\perp,i}$, in a long chain, 
corresponds to a hard subcollision between partons with momenta 
$k_{i-1}$ and $-k_{i+1}$.  $\xi{i}$ and  $x_{i}^{(-)}$ are defined 
as  $\xi{i} \equiv k_{+,i} / P_{+,tot}$ and 
$x_{i}^{(-)} \equiv - k_{-,i} / P_{-,tot}$ respectively.}
\end{figure}

A long chain as in Fig.~\ref{fig:long-chain}, with a soft probe 
($Q^2 \approx Q_0^2$) and with a local maximum 
$k_{\perp,max} = k_{\perp,i}$, corresponds to a hard scattering 
between two partons with momenta $k_{i-1}$ and $-k_{i+1}$. Due to 
the symmetry of the expression in eq.~(\ref{fq}), these two pieces 
of the chain can be interpreted as evolution from both ends towards
the central hard scattering. The momentum transfer in this hard 
subcollision is given by $k_i$, and there is a minus sign
in the momentum of one of the colliding partons, as this part of 
the chain is regarded as evolving from the top of the fan diagram, 
downwards towards $-k_{i+1}$.

\begin{figure}[t]
\begin{center}
\begin{picture}(90,90)(0,0)
\Line(0,60)(20,60)
\Text(7,53)[]{$\ki{b}$}
\Line(20,60)(25,70)
\Vertex(25,70){2}
\Text(23,78)[]{$\qi{b}$}
\Line(25,70)(50,70)
\Text(40,63)[]{$k$}
\Vertex(50,70){2}
\Text(54,78)[]{$\qi{a}$}
\Line(50,70)(70,30)
\Line(70,30)(90,30)
\Text(85,23)[]{$\ki{a}$}
\Text(40,0)[]{(a)}
\end{picture}
\hspace{2cm}
\begin{picture}(90,90)(0,0)
\Line(0,60)(20,60)
\Text(7,53)[]{$\ki{b}$}
\Vertex(20,60){2}
\Text(17,68)[]{$\qi{b}$}
\Line(20,60)(25,50)
\Line(25,50)(60,50)
\Text(40,43)[]{$k$}
\Vertex(60,50){2}
\Text(63,58)[]{$\qi{a}$}
\Line(60,50)(70,30)
\Line(70,30)(90,30)
\Text(85,23)[]{$\ki{a}$}
\Text(40,0)[]{(b)}
\end{picture}
\hspace{2cm}
\begin{picture}(90,90)(0,0)
\Line(0,60)(20,60)
\Text(7,53)[]{$\ki{b}$}
\Vertex(20,60){2}
\Text(17,68)[]{$\qi{b}$}
\Line(20,60)(40,20)
\Line(40,20)(65,20)
\Text(51,14)[]{$k$}
\Line(65,20)(70,30)
\Vertex(70,30){2}
\Text(73,38)[]{$\qi{a}$}
\Line(70,30)(90,30)
\Text(85,23)[]{$\ki{a}$}
\Text(39,0)[]{(c)}
\end{picture}
\end{center}
\caption{ \label{fig:3cases}\em The three different possibilities 
for a link.}
\end{figure}
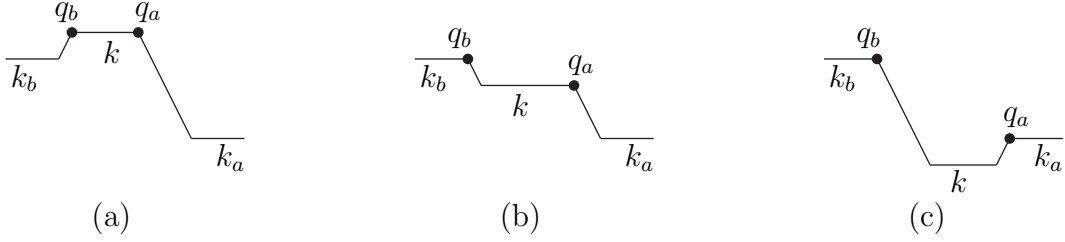

A single chain may also have more than one local maximum, which 
corresponds to two or more correlated hard subcollisions. If we 
want to obtain the inclusive jet cross section, or the total 
$E_\perp$-flow, we have to include all produced partons $q_j$ in 
the chain. To calculate this we start by studying a single 
link in the central part of a long chain. There are three different
types of links. Besides a local maximum as in 
Fig.~\ref{fig:long-chain} and Fig.~\ref{fig:3cases}a, we have also 
the two possibilities shown in Fig.~\ref{fig:3cases}b and 
Fig.~\ref{fig:3cases}c. To simplify the notation we call the gluons
$k_a$, $k$, and  $k_b$ as indicated in Fig.~\ref{fig:3cases}, and 
study the three cases:
\begin{enumerate}
\item $k_\perp > \kTi{a},\kTi{b}$. Transverse momentum conservation
implies that $\qTi{a} \approx \qTi{b} \approx \kT$, and it 
corresponds to a normal hard Rutherford scattering between two 
partons. The two colliding partons have virtualities given by 
$k_{\perp,a}^2$ and $k_{\perp,b}^2$, which both are small compared 
to the exchanged momentum $k_\perp^2$. From 
eqs.~(\ref{fk}-\ref{step-down}) we see that the link is associated 
with the following weight factor
\begin{equation}
\frac{1}{\kTpot{4}} \cdot \aspot{2}(\kTpot{2}),
\label{Rutherford}
\end{equation}
which corresponds to the cross section for gluon exchange between 
two quasireal particles.
\item $\kTi{b} > \kT > \kTi{a}$, which implies that $\qTi{a} 
\approx \kT$ and $\qTi{b} \approx \kTi{b}$. From 
eq.~(\ref{step-up}) we see that instead of the expression in 
eq.~(\ref{Rutherford}), this link is associated with the following 
weight
\begin{equation}
\frac{1}{\kTpot{2} \cdot \kTipot{b}{2}} \cdot \as (\kTpot{2}) \cdot 
\as (\kTipot{b}{2}).
\label{middle}
\end{equation}
A similar result is obtained for  $\kTi{a} > \kT > \kTi{b}$.
\item $\kT < \kTi{a},\kTi{b}$, in which case $\qTi{a} \approx 
\kTi{a}$ and $\qTi{b} \approx \kTi{b}$. The corresponding weight is
now given by
\begin{equation}
\frac{1}{\kTipot{a}{2} \cdot \kTipot{b}{2}}  \cdot \as 
(\kTipot{a}{2}) 
\cdot \as (\kTipot{b}{2}).
\label{antiRutherford}
\end{equation}
\end{enumerate}
We see that when $\kT$ is less than $\kTi{a}$ or $\kTi{b}$, one 
factor $1/ \kTpot{2}$ is replaced by $1/ \kTipot{a}{2}$ or 
$1/ \kTipot{b}{2}$ respectively. As will be discussed more in the 
following, this implies that the inclusive cross section becomes 
non-singular for small transverse momenta. (As discussed after 
eq.~(\ref{relation-new}) non-asymptotic contributions to these 
weights can be included most easily using the MC.)
%
%

Keeping $k_a$, $k$, and $k_b$ fixed, we note that integrating and 
summing over all partons to the right of the link corresponds 
exactly to the non-integrated structure function 
$\mathcal{F}(x_a,k_{\perp,a}^{2})$, where $x_a$ is the 
Bjorken variable, i. e. the scaled positive lightcone momentum, 
$x_a=k_{+,a}/P_{+,tot}$. Similarly, integration and summation of 
partons to the left of the link gives 
$\mathcal{F}(x_b,k_{\perp,b}^2)$, with $x_b$ the 
corresponding scaled negative lightcone momentum 
$x_b=x_b^{(-)}=-k_{-,b}/P_{-,tot}$. 

The inclusive jet distribution can then be written in the form 
($\shat$ is the Mandelstam variable for the hard subcollision)
\begin{eqnarray}
\frac{d \sigma_{incl}}{d \qTpot{2} dy} \propto 
\int \frac{d\xi{a}}{\xi{a}} \cdot  \frac{d\xi{b}}{\xi{b}} \cdot 
\frac { d\kTipot{a}{2}}{\kTipot{a}{2}} 
\cdot \frac {d\kTipot{b}{2}}{\kTipot{b}{2}} 
\cdot \mathcal{F}(\xi{a}, \kTipot{a}{2}) \cdot \mathcal{F}(\xi{b}, 
\kTipot{b}{2}) \nonumber \\ 
\cdot \frac{1}{2} \cdot \frac{ d \sigmahat}{d \qTpot{2}} 
(\qTpot{2}, \kTipot{a}{2}, \kTipot{b}{2}, \shat = \xi{a} \xi{b} s) 
\cdot 
\delta(y - \frac{1}{2} \ln\frac{x_a}{x_b}).
\label{incljetdistr} 
\end{eqnarray} 
In this formalism each produced parton is counted in two different 
links, one to the left and one to the right. In our notation 
$\frac{d \sigmahat}{d \qTpot{2}}$ is the inclusive cross section 
counting both emitted partons, and therefore a factor $\frac{1}{2}$
is needed to avoid double counting. The cross section 
$\frac{ d \sigmahat}{d \qTpot{2}}$ is obtained by integrating 
eqs.~(\ref{Rutherford}-\ref{antiRutherford}) over $k_\perp^2$ with 
the constraint $k_\perp^2 < \shat$. In case 1 above we will then 
get the integral (for $k_{\perp,b} > k_{\perp,a}$)
\begin{eqnarray}
\int_{k_{\perp,b}^2}^{\shat} d k_\perp^2 \cdot \frac{1}{k_\perp^4} 
\cdot 2  
\cdot \delta(q_\perp^2 - k_\perp^2) = \frac{2}{\qTpot{4}} \cdot
\theta (\qTpot{2} - \kTipot{b}{2}) \cdot \theta (\shat - \qTpot{2}).
\label{term1}
\end{eqnarray}
Here the factor 2 in front of the $\delta$-function originates from
the fact that we count both $\qTi{a}^2$ and $\qTi{b}^2$, each being
approximately equal to $\kT^2$. The contributions from cases 2 and 
3 can be calculated in the same way. The result obtained for a 
running coupling is presented in the appendix. The qualitative 
features can more easily be understood from the result obtained for
a constant $\alpha_s$, in which case we find 
(again for $k_{\perp,b} > k_{\perp,a}$)
\begin{eqnarray}
\frac{d \sigmahat}{d \qTpot{2}} &\propto& \alpha_s^2 
\left\{ \frac{2}
{\qTpot{4}}\cdot
\theta (\qTpot{2} - \kTipot{b}{2}) \cdot \theta (\shat - \qTpot{2})
+ \right. 
\nonumber \\
 &+&  \frac{1}{\qTpot{2} \cdot \kTipot{b}{2}} \cdot \theta 
(\qTpot{2} - 
\kTipot{a}{2})
\cdot \theta (\kTipot{b}{2} - \qTpot{2}) + \delta (\qTpot{2} - 
\kTipot{b}{2}) 
\cdot \frac{1}{\kTipot{b}{2}} \cdot
\ln \frac{\kTipot{b}{2}}{\kTipot{a}{2}} + \nonumber \\
&+&  \left[ \delta (\qTpot{2} - \kTipot{b}{2}) + \delta (\qTpot{2} -
\kTipot{a}{2} ) \right] \cdot
\left. \left[ \frac{1}{\kTipot{b}{2}} -  \max \left( 
\frac{1}{\shat}\,\,,\, 
\frac{Q_0^2}{k_{\perp,a}^2 k_{\perp,b}^2} \right) \right] \right\}.
\label{subcoll-cross-sect}
\end{eqnarray}
Here the first line corresponds to a ``normal'' hard subcollision, 
case 1 above. The second line corresponds to case 2, where one of 
the colliding partons has a virtuality larger than the exchanged 
momentum $-\hat{t}=k_\perp^2$, and the third line to case 3 where 
both initial partons have high virtualities. Note that if we have 
a soft cutoff $Q_0^2$, the lower limit in the $k_\perp^2$-integral 
for case 3 is given by 
$\max(\frac{k_{\perp,a}^2 k_{\perp,b}^2}{\hat{s}},Q_0^2)$.

The inclusive jet cross section is now readily obtained from 
inserting eq.~(\ref{subcoll-cross-sect}) into 
eq.~(\ref{incljetdistr}) and integrating with respect to 
$\xi{a}, \xi{b}, \kTi{a}$ and $\kTi{b}$. Due to the factorized form
$\mathcal{F}(x, \kappa) \approx x^{-\lambda}\! \cdot \!f(\kappa)$ 
in eq.~(\ref{fact}), with a power-like dependence on $x$, the 
result is independent of the rapidity of the pair, 
$y=\frac{1}{2} \ln \frac{x_a}{x_b}$, 
for fixed value of $x_a\!\!\cdot\!x_b = \shat / s$.  Thus the 
distribution corresponds to a central plateau with a height 
proportional to $(x_1\!\cdot\!x_2)^{-\lambda}$, which for fixed 
$\shat$ grows with energy proportional to $s^\lambda $.

The integration is straight forward if we assume the power-like 
approximation for $f(\kappa)$ in 
eqs.~(\ref{fact}-\ref{asymptotic}), 
$\mathcal{F}(x, \kappa) \approx x^{-\lambda} \cdot f(\kappa)
 \propto x^{-\lambda} \kappa^{\frac{\alpha_0}{\lambda} - 1} $. 
As an example, for the term in eq.~(\ref{term1}) (the first term 
in eq.~(\ref{subcoll-cross-sect})) the integral over $x_a$ and 
$x_b$ gives
\begin{equation}
\int \frac{dx_a}{x_a} \frac{dx_b}{x_b} \delta(y - \frac{1}{2} 
\ln\frac{x_a}{x_b}) \cdot \frac{1}{(x_a x_b)^\lambda} = 
\int_{q_\perp^2} 
\frac {d \hat{s}}{\hat{s}} \left(\frac{s}{\hat{s}}\right)^\lambda = 
\frac{1}{\lambda}\left(\frac{s}{q_\perp^2}\right)^\lambda
\label{s-int}
\end{equation}
while the integral over $\kTi{a}$ and $\kTi{b}$ gives (including 
also the symmetric case $k_{\perp,a}^2 > k_{\perp,b}^2$)
\begin{equation}
\alpha_s^2(q_\perp^2) \int^{\ln q_\perp^2} d \kappa_b \cdot 
\kappa_b^{\frac{\alpha_0}{\lambda}-1}  \int^{\ln q_\perp^2} d 
\kappa_a \cdot 
\kappa_a^{\frac{\alpha_0}{\lambda}-1} 
= \alpha_s^2(q_\perp^2)  \cdot \left[\frac{\lambda}{\alpha_0}
\kappa^{\frac{\alpha_0}{\lambda}}\right]^2.
\end{equation}
The full result for a running coupling, presented in the appendix, 
is rather lengthy and we write it in the form
\begin{equation}
\frac{d\sigma_{incl}^{jet}}{d\qTpot{2} dy} \propto 
\frac{s^\lambda}{\qTpot{4+2\lambda}} \cdot 
\alpha_s^2(q_\perp^2) \cdot h(\qTpot{2}).
\label{our-res}
\end{equation} 
The factor $s^\lambda/\qT^{2\lambda}$ is a consequence of the 
increase of $\mathcal{F} \sim x^{-\lambda}$ for small $x$ 
(cf eq.~(\ref{fact})), and we have also extracted a factor 
$\alpha_s^2(q_\perp^2)$. Thus the function $h(\qTpot{2})$ is 
defined in such a way, that it would be constant in the 
unrealistic case, where the parton flux is given by a scaling 
function $F(x) \sim 1/x^\lambda$, which is independent of $Q^2$. 
The result in eq.~(\ref{our-res}) is proportional to $s^\lambda$, 
but we emphasize that this is the growth of the cross section for 
jet production, and not the number of jets per event.
 
\subsection*{Results}

For a quantitative estimate of the jet cross section and a 
comparison with the ''naive'' approach, it would be suitable to use
a Monte Carlo, which can take into account also effects from 
non-asymptotic energy, non-leading terms in the parton-parton 
cross section and contributions from quarks. However, we believe 
that the qualitative features can be understood from an approximate
analytic calculation, provided the same approximations are used in 
both approaches. Thus for $F(x,q_\perp^2)$ and 
$\mathcal{F}(x,k_\perp^2)$ we use the relations in 
eq.~(\ref{relation-new}) and eqs.~(\ref{fact}-\ref{asymptotic}). 
We study purely gluonic chains and for the subcollision we use the 
leading expression 
$\frac{d\sigmahat}{d q_\perp^2} \propto 
\frac{\alpha_s^2(q_\perp^2)}{q_\perp^4}$. For the parameters 
$\lambda$ and $Q_0$ (the cut-off for small $k_\perp$) we use the 
values 0.3 and 0.6 GeV/c respectively, obtained from the MC fit to 
HERA data. Since we study purely gluonic chains, we may take 
$n_{f}=0$, which implies that the power 
$\frac{\alpha_0}{\lambda} - 1$ in eq.~(\ref{asymptotic}) is close 
to 2. The result, presented in the appendix, can be written in 
terms of simple elementary functions when this power is an integer.
Numerical calculations show that the result is insensitive to the 
exact value of the power, and that the result using 
$\frac{\alpha_0}{\lambda} - 1 \approx 2$ agrees well with results 
from the MC. For these reasons we use this value in the results 
below, and thus assume the following form:
\begin{equation}
{\cal F}(x,\kappa) \propto x^{-\lambda} \cdot \kappa^2
\label{calFapprox}
\end{equation}
which implies that the integrated structure function has the form
\begin{equation}
F(x,\kappa=\ln Q^2) \propto x^{-\lambda} \cdot \left[ \frac{1}{3}
(\kappa^3 - \kappa_0^3) + \frac{\kappa}{(\lambda+1)^2} + 
\frac{\kappa^2}{(\lambda+1)} \right]
\label{Fapprox}
\end{equation}
(Note that with our convention $F(x)$ describes the density in 
$\ln x$, and thus corresponds to $F_2$. Since we have a hard 
interaction probe we have included the running of $\alpha_s$ in 
the final vertex, which implies an extra factor 
$\ln Q^2/\ln k_\perp^2$ in the last term in 
eq.~(\ref{relation-new}).)

\begin{figure}[htb]
\mbox{\epsfig{file=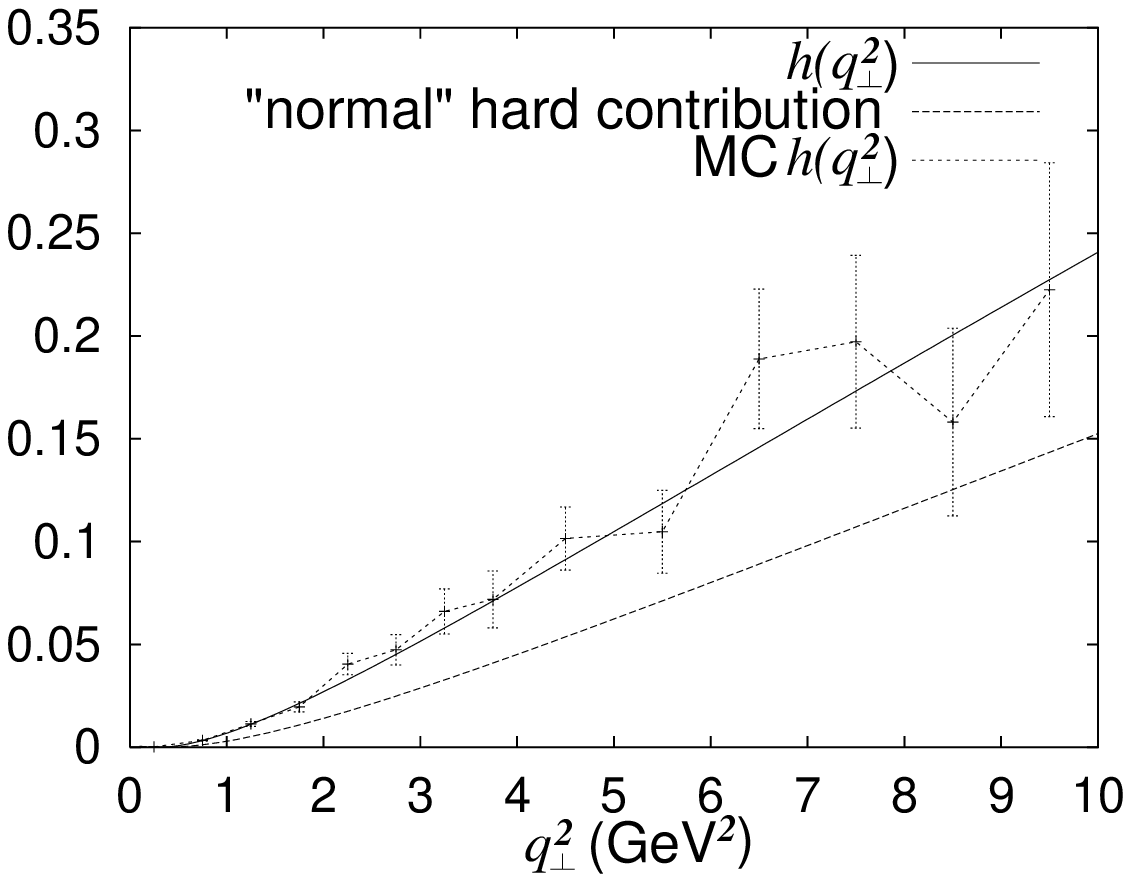,width=8.2cm}}
\mbox{\epsfig{file=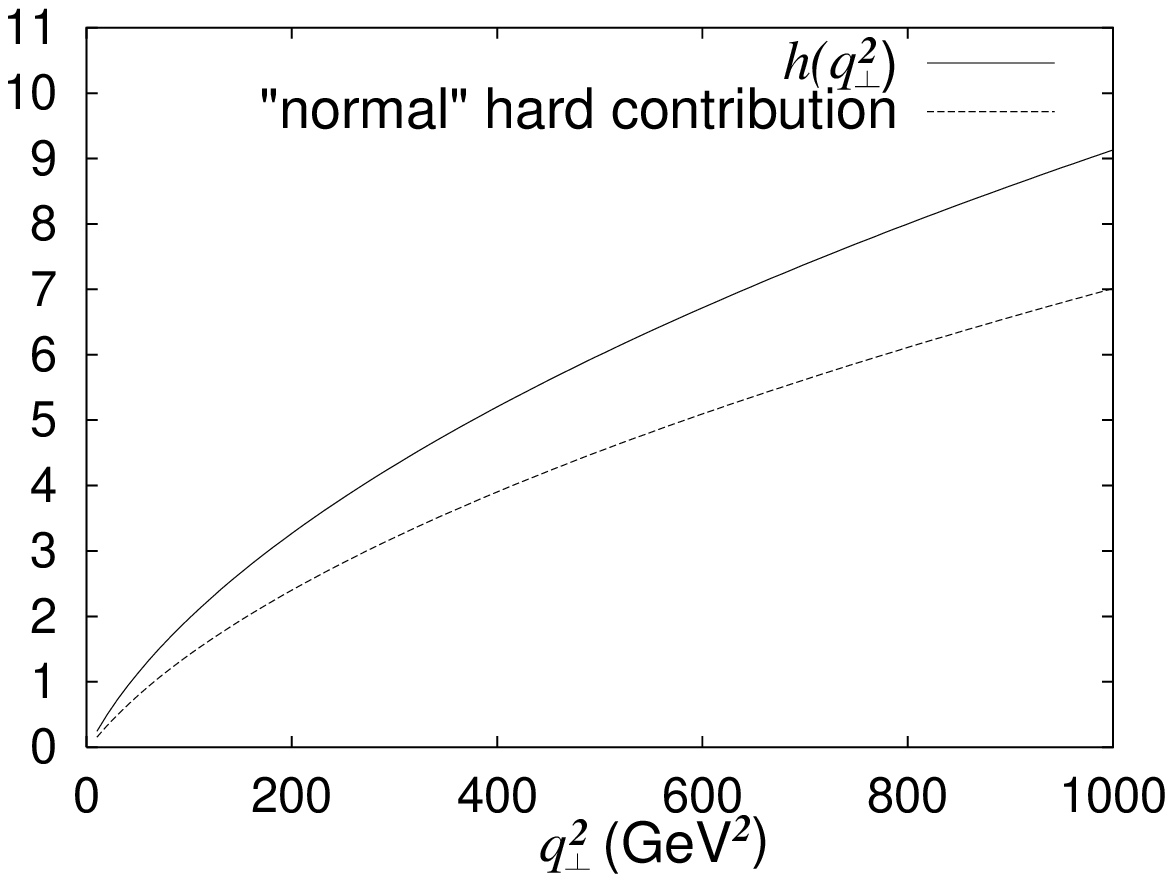,width=8.2cm}}
{ \mbox{ \begin{picture}(400,20)(0,0)\Text(123,2)[]{(a)}
\Text(352,2)[]{(b)} \end{picture} } }
\caption{  \label{fig:h} \em The transverse momentum distribution 
of $h(\qTpot{2})$ for (a) $\qTpot{2} < 10~GeV^{2}$ and 
(b)$\qTpot{2} < 1000~GeV^{2}$ according to analytical calculations 
(continuous curves) and MC simulation (short-dashed line).
(The error bars are the estimated statistical uncertainties.) 
Also shown is the contribution from ``normal hard subcollisions'' 
(long-dashed line).  The scale is arbitrary.}
\end{figure}

The transverse momentum dependence of the function $h(\qTpot{2})$ 
for a running $\alpha_s$ is shown by the solid line in 
Fig.~\ref{fig:h}. Also shown in this figure is the 
contribution from ``normal hard subcollisions'', i. e. partons 
associated to a local maximum in the $k_\perp$ chain. We see that 
this contribution dominates for large $q_\perp^2$, while for 
smaller $q_\perp^2$, the partons which are not associated to a 
local maximum play a more important role, contributing 
more than 40$\%$ for $q_{\perp}^2 < 5 \, {\mathrm {GeV}^2}$. 
For large $q_\perp^2$, $h(\qTpot{2})$ behaves as a power of 
$\ln{\qTpot{2}}$, which corresponds to the scaling violation in the
structure functions. For smaller $\qTpot{2}$, $h(q_\perp^2)$ is 
essentially linear in $\qT^2$. This implies that the total 
transverse energy flow
$d\eT/dy = \frac{1}{\sigma_{tot}}\cdot \int 
\frac{d\sigma}{d\qTpot{2} dy} d\qTpot{2} \cdot q_\perp$
is convergent for small $q_\perp$ (for $\lambda<0.5$). 
Consequently the $\eT$ distribution is limited also without a low 
$q_\perp$ cutoff. In Fig.~\ref{fig:h}a we also show the 
$q_\perp$-dependence obtained from the MC (normalized to the same 
total flow), and, as mentioned above, we see a very nice agreement 
with our analytic result.

We note that although ``normal hard collisions'' dominate for large
$q_\perp$, the other contributions correspond to approximately 
25$\%$ even for $q_{\perp}^2 \approx 1000 \, {\mathrm {GeV}^2}$. 
We expect that most of this difference would be accounted for in a 
calculation where hard collisions are calculated to next-to-leading
order, including also $2 \rightarrow 3$ parton reactions, which 
thus can take into account one extra parton. This approach would, 
however not solve the problems encountered for small or medium 
$q_\perp$.

In the literature the inclusive jet cross section is often 
estimated from a product of two (integrated) structure functions 
$F(x,q_\perp^2)$ and a hard subcollision cross section 
$\frac{d \sigmahat}{d q_\perp^2}$ for quasireal 
colliding partons. This estimate can be written in the form
\begin{eqnarray}
\frac{d\sigma_{incl}^{jet}}{d\qTpot{2} dy} &\propto& \int 
\frac{dx_a}{x_a} 
\frac{dx_b}{x_b} \delta(y - \frac{1}{2} \ln\frac{x_a}{x_b}) 
\cdot F(\xi{a}, 
\qTpot{2}) \cdot F(\xi{b}, \qTpot{2})
\cdot \frac{1}{\qTpot{4}} \cdot \alpha_s^2(\qTpot{2}) \propto 
\nonumber \\
&\propto& \frac{1}{\qTpot{4+2\lambda}} \cdot 
\alpha_s^2(\qTpot{2}) \cdot h_{naive}(\qTpot{2}).
\label{naive-res}
\end{eqnarray}
Here the function $h_{naive}$ is defined analogously to the 
function $h$ in eq.~(\ref{our-res}).

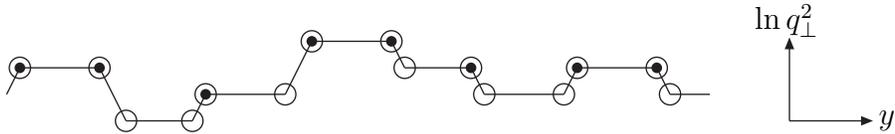
\begin{figure}[h]
\begin{center}
\scalebox{1}{\mbox{
\begin{picture}(360,50)(0,0)
\Line(15,20)(20,30)
\Line(20,30)(50,30)
\Line(50,30)(60,10)
\Line(60,10)(85,10)
\Line(85,10)(90,20)
\Line(90,20)(120,20)
\Line(120,20)(130,40)
\Line(130,40)(160,40)
\Line(160,40)(165,30)
\Line(165,30)(190,30)
\Line(190,30)(195,20)
\Line(195,20)(225,20)
\Line(225,20)(230,30)
\Line(230,30)(260,30)
\Line(260,30)(265,20)
\Line(265,20)(280,20)
\Vertex(20,30){2}
\Vertex(50,30){2}
\Vertex(90,20){2}
\Vertex(130,40){2}
\Vertex(160,40){2}
\Vertex(190,30){2}
\Vertex(230,30){2}
\Vertex(260,30){2}
\Oval(20,30)(4,4)(0)
\Oval(50,30)(4,4)(0)
\Oval(60,10)(4,4)(0)
\Oval(85,10)(4,4)(0)
\Oval(90,20)(4,4)(0)
\Oval(120,20)(4,4)(0)
\Oval(130,40)(4,4)(0)
\Oval(160,40)(4,4)(0)
\Oval(165,30)(4,4)(0)
\Oval(190,30)(4,4)(0)
\Oval(195,20)(4,4)(0)
\Oval(225,20)(4,4)(0)
\Oval(230,30)(4,4)(0)
\Oval(260,30)(4,4)(0)
\Oval(265,20)(4,4)(0)
\LongArrow(310,10)(340,10)
\LongArrow(310,10)(310,40)
\Text(310,48)[]{$\ln \qTpot{2}$}
\Text(348,10)[]{$y$}
\end{picture} }}
\end{center}
\caption{ \label{fig:comparison}\em The outgoing partons according 
to a correct formalism (dots) and according to a naive approach 
(circles).}
\end{figure}

It is interesting to compare this ``naive'' estimate with our 
result in eq.~(\ref{our-res}). The relation between 
$F(x,q_\perp^2)$ and the non-integrated structure function 
$\mathcal{F}(x,k_\perp^2)$ is shown in eq.~(\ref{relation-new}). 
There are two contributions to $F$, one (ordered) contribution 
with $q_\perp > k_\perp$, and another suppressed contribution 
with $q_\perp < k_\perp$. When one or both colliding partons are 
virtual, there is a suppression factor $q_\perp^2/k_{\perp,a}^2$ 
and/or $q_\perp^2/k_{\perp,b}^2$. With a cross section assumed to 
be proportional to $\frac{\alpha_s^2(q_\perp^2)}{q_\perp^4}$ as in 
eq.~(\ref{our-res}), we see that each link is given just the weight
presented in eq.~(\ref{Rutherford}), eq.~(\ref{middle}), or 
eq.~(\ref{antiRutherford}). The difference lies in the fact that in
eq.~(\ref{naive-res}) both outgoing partons are assumed to have 
transverse momenta given by the momentum transfer in the collision.
This underestimates the $q_\perp$ for collisions corresponding to 
the links in Fig.~\ref{fig:3cases}b and Fig.~\ref{fig:3cases}c. 
This underestimate is compensated by the fact that 
every link in the fan diagram corresponds to a hard collision in 
eq.~(\ref{naive-res}). Thus this equation corresponds to two 
produced partons per link instead of one. This double counting does
not give a full factor of two, however, due to the underestimate of
the $q_\perp$ mentioned above. The result is illustrated in 
Fig.~\ref{fig:comparison}. The really emitted partons 
in a chain are marked by a dot, while the naive expression in 
eq.~(\ref{naive-res}) corresponds to a final state parton at every 
point marked by a circle. Thus in the naive calculation only the 
partons at a local maximum are given their proper weight. 
Those at a minimum value for $k_\perp$ should not be included 
at all, as no real emitted partons have these $k_\perp$-values, 
and those in-between are counted twice.
%

\begin{figure}[t]
\begin{center}
\mbox{\epsfig{file=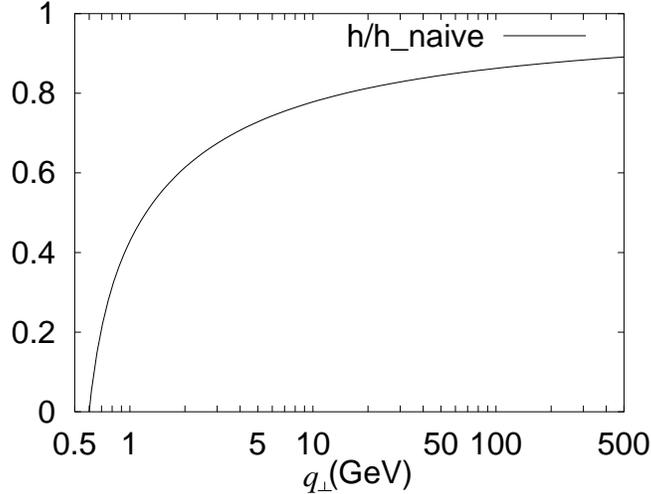,width=9.5cm}}
\end{center}
\caption {\label{fig:h/hnaive} \em The ratio 
$h(\qTpot{2}) / h_{naive}(\qTpot{2})$ between the cross section in 
our approach and the naive expectation in eq.~(\ref{naive-res}), 
as a function of $\qT$. (Note the logarithmic scale on the x-axis.)}
\end{figure}

\begin{figure}[h]
\begin{center}
\mbox{\epsfig{file=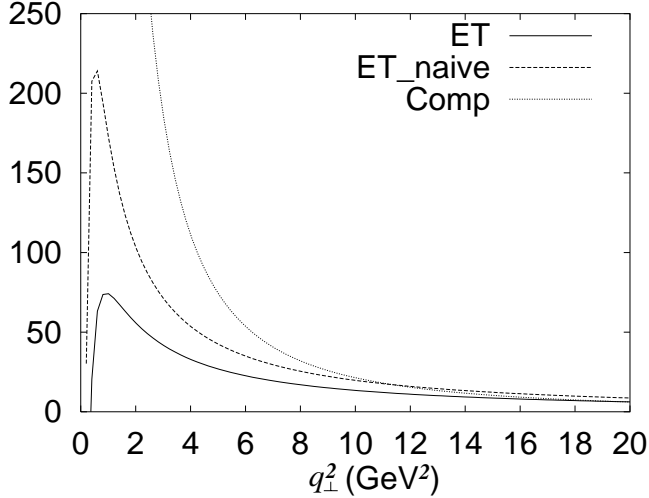,width=9.5cm}}
\end{center}
\caption{\textit{ Transverse energy distribution, 
$d \eT / dy d \qTpot{2}$, according to our calculations 
(continuous curve) and the ''naive'' estimate in 
eq.~(\ref{naive-res}) (dashed curve). For comparison we also show 
the result from scaling structure functions, $F(x)$ independent of 
$\qTpot{2}$ (dotted curve). The normalization is arbitrary.} 
\label{fig:ET} }
\end{figure}

\begin{figure}[th]
\begin{center}
\mbox{\epsfig{file=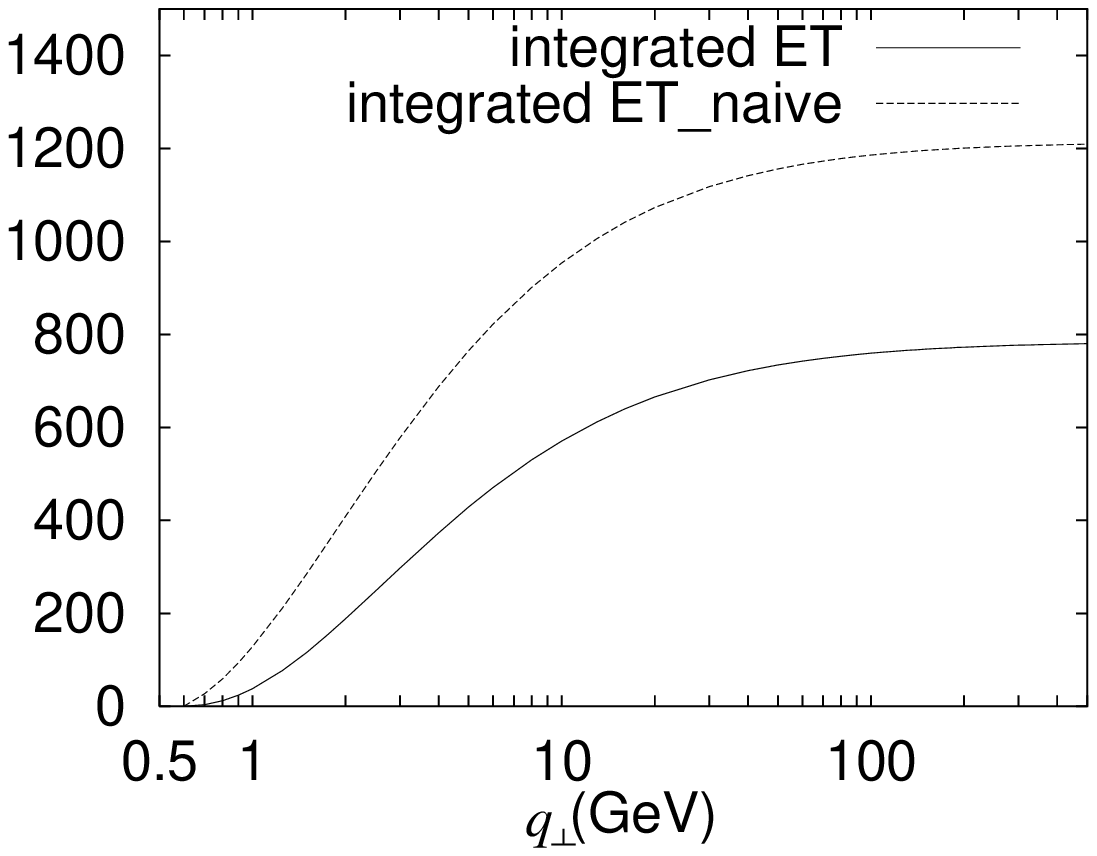,width=9.5cm}}
\end{center}
\caption{\textit{ Integrated transverse energy distribution, 
$\int^{q_\perp^2} dq_\perp^2 d E_\perp / dy dq_\perp^2$, according 
to our calculations (continuous curve) and the ''naive'' estimate 
in eq.~(\ref{naive-res}) (dashed curve). The normalization is as in 
Fig.~\ref{fig:ET}.} \label{fig:intET} }
\end{figure}

Fig.~\ref{fig:h/hnaive} shows the ratio 
$h(\qTpot{2}) / h_{naive}(\qTpot{2})$ 
as a function of the transverse momentum (note the logarithmic 
$q_\perp$-scale). For large $\qT$ the Rutherford contribution 
dominates in both cases, making 
$h(\qTpot{2}) \approx h_{naive}(\qTpot{2})$. For smaller 
$\qT$, on the other hand, the naive expectation gives a significant 
overestimate. This can also be concluded from Fig.~\ref{fig:ET},
which shows the distribution in transverse energy, 
$d \eT / dy d \qTpot{2} = \qT \cdot dn / dy d \qTpot{2}$. We see 
that the total $\eT$ in the minijet region,
 $q_\perp \stackrel{<}{\sim}$ 5 GeV, is almost a factor 2 larger 
in the naive estimate. It is often realized that this large 
contribution for very low $q_\perp$ must be unphysical, 
and therefore a soft cutoff, $q_{\perp,min}$, is introduced in many 
calculations. This cutoff is often assumed to be around $2$ GeV and
slowly growing with energy (see e.g. \cite{cutoff}). In 
Fig~\ref{fig:intET} we show the integrated transverse energy 
$\int^{q_\perp^2} dq_\perp^2 \frac{d E_\perp}{dy dq_\perp^2}$. 
From this figure we see that 
\begin{equation}
\int_{Q_0^2}^\infty dq_\perp^2 \left. \frac{d E_\perp}
{dy dq_\perp^2}
\right|_{our\,result}  \approx \int_{q_{\perp,min}^2}^\infty 
dq_\perp^2 \left. 
\frac{d E_\perp}{dy dq_\perp^2}\right|_{naive}
\end{equation}
for $q_{\perp,min} \approx 2.1$ GeV. (Remember that the low 
$k_\perp$-cutoff $Q_0$ was given by 0.6 GeV.) This means that for 
asymptotic energies
%
%
the total $E_\perp$ in our approach equals the naive result 
when $q_{\perp,min} \approx 2.1$ GeV. For smaller energies the 
maximal value of $q_\perp$ is limited, which implies that a 
somewhat smaller value of $q_{\perp,min}$ is needed. A reliable 
quantitative estimate would need a calculation which includes quark
jets and non-leading contributions, which could be obtained with 
the help of the MC program. We note in particular that in our 
approach the corresponding effective cutoff of the naive approach, 
$q_{\perp,min}$, saturates for very large energies. In the 
conventional approach it is difficult to make predictions for 
higher energies without a physical understanding of the energy 
dependence of $q_{\perp,min}$.

Only for comparison, we have in Fig.~\ref{fig:ET} also included the 
distribution one would obtain from eq.~(\ref{our-res}) if the 
structure function $F$ were scaling and independent of $q_\perp^2$.
(The normalization is adjusted to our result for large momenta.) 
As is seen, this would give a very much larger (and totally 
unrealistic) increase for small $q_\perp$.  

We end this section by noting that one chain forms a set of 
correlated jets. The fact that the $k_\perp$-distribution saturates
for long chains (cf eq.~(\ref{fact})) implies that not only the 
$p_\perp$ of the jets, but also their density (the number of jets 
per unit rapidity) is independent of the length of the chain, i. e.
independent of the energy in the collision. The jet density grows 
because the number of possible chains increases proportional to 
$s^\lambda$. It is also conceivable that the number of chains 
is not random, but e. g. correlated with the impact parameter, 
so that central collisions have more and peripheral collisions 
fewer chains \cite{correlations}. In this paper we have only 
studied the average jet multiplicity, and we postpone the study of 
different types of correlations to future investigations.

The possible correlation between different chains may also affect 
the total cross section. We want to stress that the results in 
eq.~(\ref{our-res}) correspond to the cross section for jet 
production. To find the number of jets per event we have to divide 
by the total cross section, which is not directly obtained from our
formulae.

\section{Conclusions}

A good understanding of jet and minijet production, and transverse 
energy flow, is essential for a proper interpretation of new 
phenomena in $pp$ collisions at LHC or a possible formation of a 
quark-gluon plasma in nucleus collisions at RHIC or LHC. For very 
high $p_\perp$ the parton flux can be described by DGLAP 
($k_\perp$-ordered) evolution. The large momentum transfer between 
the colliding partons implies that the two evolving chains are 
independent of each other. For moderate and smaller $p_\perp$ 
(and high energies) we enter the BFKL regime and non-ordered chains
become important. Here it is essential to take coherence effects 
and correlations into account. As we can then have several hard 
subcollisions in a single chain or ladder diagram, it is also 
important to avoid double counting.

The parton evolution in the BFKL regime can be described e. g. by 
the CCFM model or the semisoft formalism. The LDC model is a 
reformulation and generalization of the CCFM formalism, and the 
symmetric structure of the LDC model makes it particularly suited 
for a description of (mini)jet production. The result can be 
interpreted as the production probability for exclusive final 
states, which makes it convenient for treating the problem of 
double counting. The (mini)jet cross section is described in a 
$k_\perp$-factorizing form in terms of non-integrated structure 
functions and off-shell subcollision cross sections. The result is 
a dynamical suppression of small-$p_\perp$ jets, which removes the 
strong sensitivity to a low-$p_\perp$ cutoff in ``naive'' estimates
based on integrated structure functions.

\section{Acknowledgments}

We want to thank Bo Andersson and Leif L\"onnblad for valuable 
discussions, and L\"onnblad for help with the Monte Carlo 
simulations.

\section{Appendix}
\label{Appendix}

In this appendix we derive the explicit form of the function
$h(\qTpot{2})$, which appears in the expression for the total cross
section in eq.~(\ref{our-res}), for the case of a running coupling 
constant $\alpha_s \propto 1/\ln(k_\perp^2/\Lambda^2) \equiv 
1/\kappa$. In each vertex the scale in  $\alpha_s$ is taken to be 
the largest transverse momentum, which due to the consistency 
constraint in eq.~(\ref{cc}) approximately coincides with the 
largest virtuality. Thus we write the cross section given by 
eq.~(\ref{incljetdistr}) in the form
\begin{equation}
\frac{d\sigma_{incl}^{jet}}{d\qTpot{2} dy} \propto 
\frac{s^\lambda}{\qTpot{4+2\lambda}} \cdot 
\alpha_s^2(q_\perp^2) \cdot h(\qTpot{2}).
\label{our-res-app}
\end{equation} 
As before, we consider separately each of the three possible cases
for the relative sizes of the transverse momenta of the gluons
$k_a$, $k$, and  $k_b$ in Fig.~\ref{fig:3cases}:\\

  Case 1. $k_\perp > \kTi{a},\kTi{b}$, which implies that 
$\qTi{a} \approx   \qTi{b} \approx \kT$,

  Case 2. $\kTi{b} > \kT > \kTi{a}$, which implies that $\qTi{a} 
\approx  \kT$ and $\qTi{b} \approx \kTi{b}$,

  Case 3. $\kT < \kTi{a},\kTi{b}$, which implies that 
$\qTi{a} \approx  \kTi{a}$ and $\qTi{b} \approx \kTi{b}$.\\

We also assume the power-like approximation 
for $f(\kappa)$ in eqs.~(\ref{fact}-\ref{asymptotic}), 
$\mathcal{F}(x, \kappa) \approx x^{-\lambda} \cdot f(\kappa)
 \propto x^{-\lambda} \kappa^{\frac{\alpha_0}{\lambda} - 1} 
\approx x^{-\lambda} \cdot \kappa^2$, where $\frac{\alpha_0}
{\lambda} - 1$ is approximately equal to 2.

We start by considering the contribution from case 1. Here the 
scale in $\alpha_s$ is given just by $k_\perp^2 \approx q_\perp^2$.
Thus the only modification to eq.~(\ref{term1}) is a factor 
$\alpha_s^2(q_\perp^2) \propto 1/\ln^2(q_\perp^2)$, which is 
factored out explicitly in the definition of $h(q_\perp^2)$ in 
eq.~(\ref{our-res-app}). Thus we find

\begin{eqnarray}
\left. \frac{d \sigmahat}{d \qTpot{2}} \right|_{contr~1}  
&\propto& 
2 \cdot \frac{1}{q_\perp^4} \cdot \frac{1}{\ln^2(q_\perp^2)} .
\end{eqnarray}

We have a factor 2 in the inclusive cross section, because we count
both outgoing partons. When this expression is inserted into 
eq.~(\ref{incljetdistr}), the integrations with respect to 
$\kTi{a}$ and $\kTi{b}$
give\\
\begin{equation}
\frac{1}{2} \int_{\kappa_{0}}^{l_q}
d \kappa_{b} \int_{\kappa_{0}}^{l_q} d\kappa_{a} \cdot 
\frac{2}{\qTpot{4}}  \cdot
\frac{1}{l_q^{2}} \cdot \kappa_{a}^{2} \cdot \kappa_{b}^{2} 
=
\frac{1}{\qTpot{4}} \cdot \frac{1}{l_{q}^{2}} \cdot \frac{1}{9} 
\cdot
(l_{q}^{3}-\kappa_{0}^{3})^2
\end{equation}
where $l_{q} \equiv \ln{\qTpot{2}}$ and 
$\kappa_{0} \equiv \ln{Q_{0}^{2}}$. Here the factor $1/2$ 
compensates for double counting as described in the main text.
The remaining integrations with respect to $x_{a}$
and $x_{b}$ (cf eq.~(\ref{s-int})) give finally:\\
\begin{equation}
\left. \frac{d \sigma}{d \qTpot{2} d y} \right|_{contr~1}
\propto
\frac{s^\lambda}{\qTpot{4+2\lambda}} \cdot 
\frac{1}{l_{q}^{2}} \cdot \frac{1}{\lambda} \cdot \frac{1}{9} \cdot 
(l_{q}^{3}-\kappa_{0}^{3})^2.
\label{case1}
\end{equation}

For case 2, the subcollision cross section 
$\frac{d \sigmahat}{d \qTpot{2}}$ is given by\\
\begin{eqnarray}
\left. \frac{d \sigmahat}{d \qTpot{2}} \right|_{contr~2}  
&\propto&  
\int_{\kTipot{a}{2}}^{\kTipot{b}{2}} d \kTpot{2} \cdot 
[ \delta (\qTpot{2}-\kTipot{b}{2}) + \delta (\qTpot{2}-\kTpot{2})] 
\cdot \frac{1}{\kTipot{b}{2}} \cdot \frac{1}{\kTpot{2}}  \cdot
\frac{1}{\kappa_b} \cdot \frac{1}{\kappa}  \nonumber \\
&=& 
\frac{1}{\kTipot{b}{2}} \cdot \frac{1}{\kappa_{b}} \cdot 
\left[\delta(\qTpot{2}-\kTipot{b}{2}) \ln(\frac{\kappa_b}
{\kappa_a}) +
\theta(\kTipot{b}{2}-\qTpot{2}) \theta(\qTpot{2}-\kTipot{a}{2}) 
\cdot
\frac{1}{q_\perp^2} \cdot \frac{1}{l_q}\right]
\end{eqnarray}
After insertion into eq.~(\ref{incljetdistr}), calculations 
analogous to the ones in case 1 give the following result for the 
contribution to the total cross section: \\
\begin{eqnarray}
\left. \frac{d \sigma}{d \qTpot{2} d y} \right|_{contr~2}  
&\propto&  
\frac{s^\lambda}{\qTpot{4+2\lambda}} \cdot 
\frac{1}{l_{q}^{2}} \cdot \frac{1}{3} \cdot 
\left\{ 
\frac{1}{\lambda} 
\left[ \frac{l_{q}^{6}}{3} - \frac{\kappa_{0}^{3} l_{q}^{3}}{3} -
\kappa_{0}^{3} l_{q}^{3} \cdot ln \left(\frac{l_{q}}{\kappa_{0}}
\right)
\right] +
\right. 
\nonumber \\
& &
\left. 
+ l_{q} \cdot (l_{q}^{3}-\kappa_{0}^{3}) \cdot
\left[ \frac{1+ l_{q}}{\lambda (\lambda+1)} - \frac{1}
{(\lambda+1)^2}
\right] 
\right\} 
\label{case2}
\end{eqnarray}

In case 3 the appropriate expression for the subcollision 
cross section $\frac{d \sigmahat}{d \qTpot{2}}$ is \\
\begin{eqnarray}
\left. \frac{d \sigmahat}{d \qTpot{2}} \right|_{case~3}  
&\propto&  
\int^{\kTipot{a}{2}} d \kTpot{2} \cdot 
[ \delta (\qTpot{2}-\kTipot{a}{2}) + \delta 
(\qTpot{2}-\kTipot{b}{2})] 
\cdot \frac{1}{\kTipot{b}{2}} \cdot \frac{1}{\kTipot{a}{2}}  \cdot
\frac{1}{\kappa_{b}} \cdot \frac{1}{\kappa_{a}},
\end{eqnarray}
where the lower integration limit is given by 
max$(\frac{\kTipot{a}{2} \kTipot{b}{2}}{\shat},Q_{0}^{2})$.
Inserting this into eq.~(\ref{incljetdistr}) and performing the 
integrations we obtain\\
\begin{eqnarray}
\left. \frac{d \sigma}{d \qTpot{2} d y} \right|_{contr~3}  
&\propto&  
\frac{s^\lambda}{\qTpot{4+2\lambda}} \cdot  
\left[
\frac{1}{\lambda(\lambda+1)} \cdot l_{q} \cdot
\left( 
\frac{l_{q}^{2} - \kappa_{0}^{2}}{2} + l_{q} - \kappa_{0}
\right) -
\right. 
\nonumber \\
& &
\left. 
\frac{1}{(\lambda+1)^2} \cdot l_{q} \cdot (l_{q} - \kappa_{0}) 
\right]. 
\label{case3}
\end{eqnarray}

Adding the three contributions, and including the symmetric 
situation $\kTi{a} > \kTi{b}$ to case 2, we find finally for the 
function $h(q_\perp^2)$ in eq.~(\ref{our-res-app})
\begin{eqnarray}
h(\qTpot{2})  
&\propto&  
\left\{ 
\frac{1}{\lambda} \cdot \frac{1}{9} \cdot 
(l_{q}^{3}-\kappa_{0}^{3})^2 
\right\} + \nonumber \\
&+&
\left\{
\frac{1}{\lambda} \cdot \frac{1}{3} \cdot
\left[ 
\frac{l_{q}^{6}}{3} - \frac{\kappa_{0}^{3} l_{q}^{3}}{3} -
\kappa_{0}^{3} l_{q}^{3} \cdot 
ln \left(\frac{l_{q}}{\kappa_{0}}\right)
\right] 
+
\frac{1}{3} \cdot l_{q} \cdot (l_{q}^{3}-\kappa_{0}^{3}) \cdot 
\left[
\frac{1+l_{q}}{\lambda(\lambda+1)} - \frac{1}{(\lambda+1)^2}
\right]
\right\} + \nonumber \\     
&+&  
\left\{
\frac{1}{\lambda(\lambda+1)} \cdot l_{q}^3 \cdot 
(\frac{l_{q}^2}{2} - \frac{\kappa_{0}^2}{2} + l_{q} - \kappa_{0}) 
- \frac{1}{(\lambda+1)^2} \cdot l_{q}^3 \cdot (l_{q}-\kappa_{0})
\right\}.
\label{...}
\end{eqnarray}
This can be compared with the ``naive'' expression obtained from 
eq.~(\ref{Fapprox}). With the same normalization we get
\begin{equation}
h_{naive} \propto \frac{2}{\lambda} \left[ \frac{1}{3}(l_q^3 - 
\kappa_0^3) + \frac{l_q}{(\lambda+1)^2} + \frac{l_q^2}{(\lambda+1)}
\right]^2
\label{hnaive},
\end{equation}
where the factor 2 in the inclusive cross section follows because 
each hard subcollision produces two jets. We see that the leading 
term for large $q_\perp$, $\frac{2}{9\lambda}(\ln q_\perp^2)^6$, 
is the same in both cases, but $h_{naive}$ is significantly larger 
for smaller $q_\perp$-values.
\end{document}